# Kinetic Inductors Enable Reversible Logic

Erik P. DeBenedictis
*Zettaflops LLC*
Albuquerque, NM, USA
erikdebenedictis@gmail.com, (ver. June 5, 2026)

***Abstract*—Reversible logic has long promised substantial reductions in energy dissipation, yet prior demonstrations have not scaled to commercially relevant systems. This work presents a quantitative framework for evaluating reversible logic through a process termed CMOS conversion, in which a conventional CMOS design is transformed into a functionally equivalent reversible implementation and compared using common performance metrics. The framework combines planning equations, kinetic-inductor energy-storage models, a four-phase 4LC energy-recycling power supply, and RLC-based simulation methods that account for data-dependent loading effects. The analysis identifies inductor loss as a fundamental limitation of conventional approaches and shows that high-energy-density kinetic inductors provide essential design margin for scaling reversible systems. Using representative device parameters, the framework suggests that selected cryogenic CMOS qubit-controller circuits could be converted to reversible logic using available or near-term technologies. Rather than claiming commercialization of reversible logic in general, the paper provides a methodology for assessing its feasibility and potential benefits across future applications.**

## I. Background

Beginning in the 1960s, physicists including Landauer [1], Bennett [2], and Feynman [3] identified the minimum heat dissipation in irreversible logic gates (e.g., AND/OR gates). In contrast, no fundamental lower bound exists for dissipation in reversible classical gates, such as the Toffoli gate. In 1984, the U.S. Government advisory group JASON recommended applying reversible logic principles to reduce the energy consumption of integrated circuits [4].

DARPA subsequently funded an "adiabatic switching" test chip to evaluate reversible logic principles when implemented via energy recycling, as illustrated in Fig. 1 [5]. Although logic operations were performed using the same nFET and pFET transistors as in conventional CMOS circuits (Fig. 1b), the resulting adiabatic circuits return most of the ½$CV^2$ switching energy to an energy-recycling power supply (Fig. 1a).

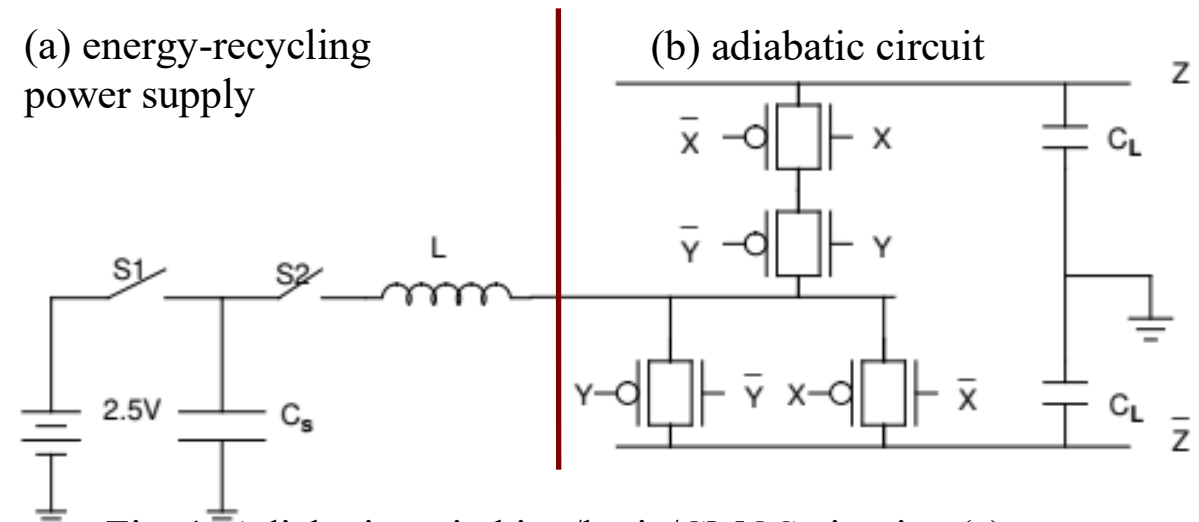

Fig. 1. Adiabatic switching/logic/CMOS circuits. (a) energy-recycling power supply. (b) adiabatic switching circuit

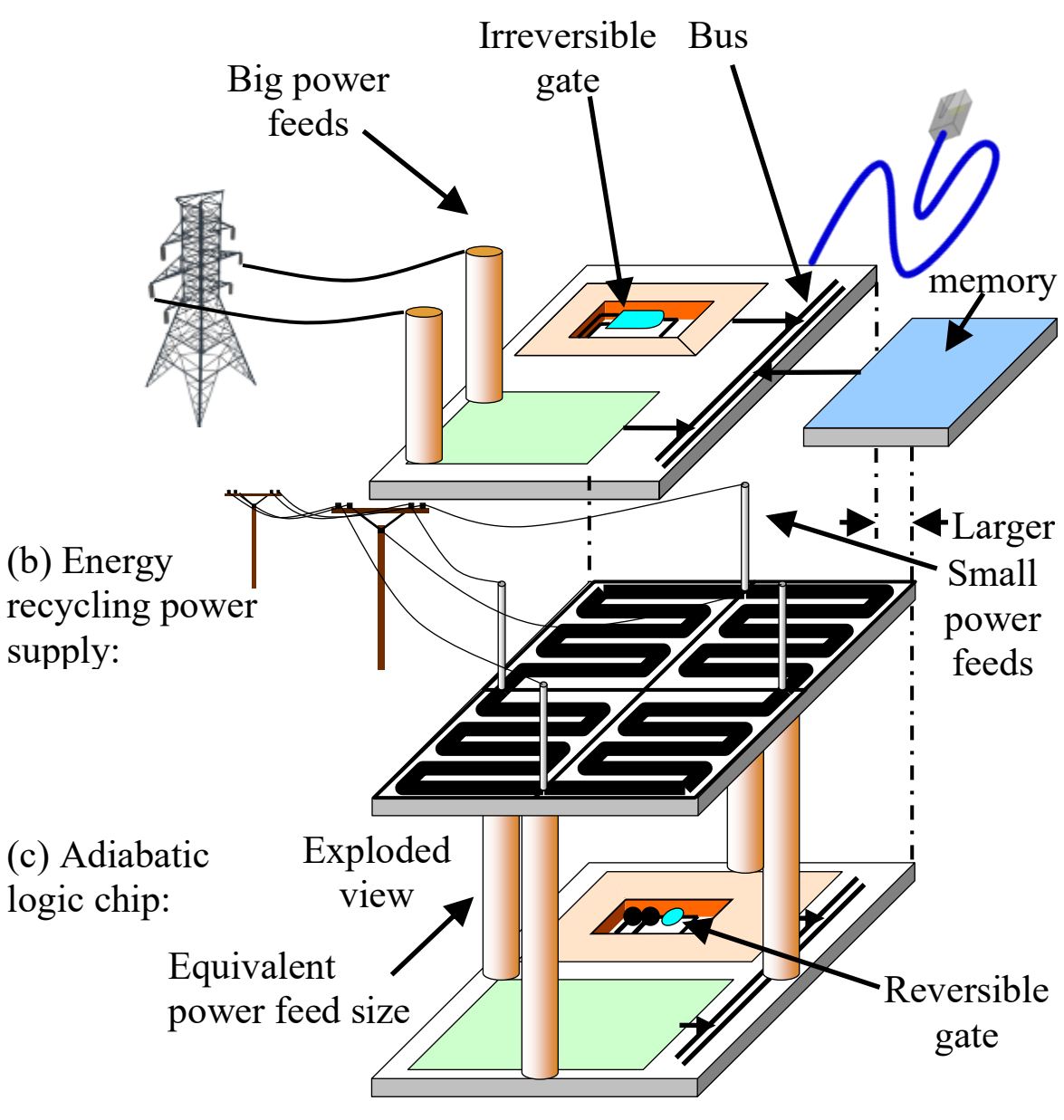

Fig. 2. CMOS conversion

The system-level objective was to convert conventional logic in a CMOS chip (Fig. 2a) into reversible logic gates (Fig. 1b and Fig. 2c) without altering functionality, thereby enabling the direct substitution of CMOS logic with lower-power reversible equivalents.

To date, about a dozen R&D programs have fabricated adiabatic CMOS test chips. While many performed as expected, none progressed to the integration complexity scaling required for commercial viability. These efforts introduced novel circuit design techniques [6, 7] and circuit families including T-gate logic (a.k.a. 2LAL) [8, 9], S2LAL [10], RERL [11], and specialized bus structures [12]. A funded startup [13] reportedly developed additional circuit techniques that remain undisclosed. (Note that these programs did not use the integrated approach illustrated in Fig. 2b; rather, [9] employed a MEMS resonator, and the rest used conventional external inductors. More recently, resonant rotary oscillators have also been investigated as a scalable alternative to LC-based power-clock generation [26].)

A key observation is that the power flux into a reversible chip is approximately equal to that of the original CMOS; the reduction occurs between the primary power source and the energy-recycling power supply. In Fig. 2, the copper-colored cylinders are sized to represent the current required for charging interconnects and gates. Because the circuits in Fig. 2a and Fig. 2c perform identical functions using the same semiconductor process, their currents are similar.

The primary obstacle to commercial viability appears to have been the Q of the inductor in Fig. 1a, which was identified as early as 1993 [27]. The fundamental question is whether a device exists with sufficient energy storage capacity at gigahertz (GHz) clock rates to match the power flux requirements.

### A. Potentially an Infinite Loop

The repeated funding of similar concepts achieving the same milestone [8–13] suggests an "infinite loop." In this case, each iteration generated arguments that reversible logic faces fundamental limitations, including control complexity, leakage currents, non-adiabatic losses, area overhead, throughput penalties, and immature design tools. However, both venture capital and government R&D funding processes involve due diligence, which is analogous to peer review. As a result, each iteration effectively transforms claimed limitations into acceptable risks for the next iteration.

Thus, the current unmet objective is not to secure funding for another test chip, but to develop a test chip that can scale to commercial viability. An R&D project satisfying the following criteria should break out of the loop:

1. It identifies a limitation present in prior work that has not been articulated in the literature.
2. It resolves the identified limitation.
3. It otherwise builds upon prior work [6–13], thus leveraging existing validation.

### B. Kinetic Inductance

In superconducting systems, inductors based on the kinetic energy of charge carriers—rather than magnetic fields—have recently become feasible due to the availability of layers of high-kinetic-inductance (HKI) material. These inductors offer two standout properties.

First, kinetic inductors offer an advantage in volume-constrained applications. Conventional magnetic inductors store energy in magnetic fields that extend in three dimensions (3D) by the diameter of the coil (spiral) or conductor spacing (meander). As inductors are placed closer to other inductors or data wires, managing crosstalk requires 3D magnetic field simulation—ultimately encountering a lower bound on system volume. Since kinetic inductors do not rely on magnetic fields for energy storage, they can be placed closer together before complex 3D simulation is needed, and they occupy a smaller footprint at this lower bound [14].

Second, high power flux is a standout property. An inductor used for energy storage is analogous to a battery, but the charge–discharge cycle time is fundamentally limited by the dimension of the inductor divided by the speed of light. For GHz charge–discharge cycles, the higher energy density of kinetic inductors [14] leads to a higher power flux. Several foundries offer HKI layers, and this document uses the SeeQC process [15] as a representative example. The process provides a niobium layer with critical current $I_c$ = 2.5 mA/μm of wire width, inductance per square $L_\square$ = 8.5 pH/□, minimum wire-to-wire spacing $s_{min}$ = 1 μm, and minimum wire width $w_{min}$ = 0.8 μm. (Note: In this context, the maximum current in a wire is defined as $I_c$ multiplied by the wire width, which diverges from the conventional definition of critical current.)

The energy stored in an inductor is given by $E = ½LI^2$. Using the SeeQC process parameters, we consider the expression $E = ½L_\square I_c^2$, which represents an idealized maximum energy capacity per unit area of an HKI sheet. Similar to $R_\square$ for resistors, this "$L_\square$ model" may represent a qualitative change for designing systems with inductors. While the use of $L_\square$ neglects overheads such as geometric manufacturing variance, electrical headroom, magnetic fields, wire spacing, turns, and contacts, it supports upfront planning by deferring the need for 3D and statistical simulation until later in the design process.

### C. Upfront Analysis of CMOS Conversion

The body of this document presents a quantitative analysis of CMOS conversion (Fig. 2) applied to chips of varying complexity using various types of inductors.

The process starts by gathering parameters on the original CMOS chip, extensions to the process design kit (PDK) for kinetic inductors, parameters for the reversible logic, and unique correction factors.

The main calculation analyzes the original CMOS to obtain the effective circuit capacitance per unit area and then applies scaling rules to the structure in Fig. 2b. This yields an inductor geometry for various inductor types with the property that its resonant frequency matches the target clock frequency of the reversible logic.

Backend calculations are then performed for each inductor type.

For non-superconducting inductors, the back-end calculations determine the resistance of the inductors and their Q factor. Because this calculation assumes the inductor is in a vacuum, the result represents an optimistic upper bound. Thus, the non-superconducting back-end calculations are helpful in understanding whether kinetic inductors are essential for a specific application.

For kinetic inductors, the power density of the reversible logic is divided by the kinetic inductors' power flux, based on either the idealized energy capacity $E = ½L_\square I_c^2$ or estimates derived from the computed inductor geometry. Rounding the quotient up to the nearest integer yields the required number of HKI layers. It then becomes the designer's responsibility to find a suitable foundry. Equivalently, if the quotient exceeds unity, a single HKI layer is insufficient.

There are some cross-cutting issues. For example, Appendix 1 describes a resonator circuit with superior properties, but resonates √2 faster than a conventional LC circuit. This document also describes infrastructure circuitry (e.g., generators for arbitrary power-clock waveforms) that would be included in the reversible redesign at the circuit level.

### D. Use Case Summaries

The body of this document contains formulas for a quantitative analysis of CMOS conversion, and these equations have been embodied in both an Excel spreadsheet and a Python script that can be found in the supplemental material. Readers can use thus perform "what if" tests and create parameter sweeps.

The body of the document and the supplemental materials include several use cases, which are summarized below:

The first use case is the Horse Ridge cryogenic controller chip [16]. Published data on the 4 × 4 mm chip indicate that its cryogenic CMOS dissipates between 10 mW and 165 mW at 4 K, depending on activity. Applying the SeeQC process parameters to an equivalent area operating at 1 GHz yields an estimated power capacity of 425 mW, indicating that a single HKI layer is sufficient. This approach is also applicable to other quantum controller chips [17].

For room-temperature applications, operation with kinetic inductors at cryogenic temperatures enables the reduction of $V_{dd}$ from 1.0 V to 0.4 V. This permits the conversion of a 14 W room-temperature chip into an approximately 1 $cm^2$ reversible logic chip operating at 1 GHz.

Some proposals suggest converting high-end AI accelerators (e.g., NVIDIA B200) to reduce kilowatt-level dissipation; however, such designs would require hundreds of HKI layers. Flash memory employs multilayer (*N*-layer) three-dimensional (3D) fabrication techniques to increase storage capacity by a factor of *N* without a proportional increase in expensive and unreliable lithographic steps [18]. Appendix 2 explains how to adapt this approach to increase the energy capacity of kinetic inductors.

### E. Arbitrary Power-Clock Generation

This document introduces a novel resonator circuit and auxiliary circuits capable of generating all possible power-clock waveforms that could result from the 4LC circuit.

### F. High-$T_c$

Suitable high-$T_c$ kinetic inductors are available [19] and are considered in this document. Interest in cryo-CMOS at 77 K is reflected in active government investment; for example, DARPA's LTLT program targets a 25× performance/power improvement for CMOS operating at liquid-nitrogen temperatures [25].

### G. Ultra-Steep-Slope Transistors

While this document primarily introduces an energy-recycling power supply (Fig. 1a), independent advances in ultra-steep-slope transistors [20] may provide synergistic benefits, particularly at millikelvin temperatures. Prior work in quantum computing includes cryogenic CMOS (cryo-CMOS) qubit controllers [16, 17]. These 4 K-class circuits have demonstrated success for small-scale systems; however, scaling challenges include heat dissipation and noise. In particular, prior work notes that "qubit … gates remain sensitive to electrical noise, arising, for instance, from volt-scale, sub-nanosecond switching of proximal CMOS transistors" [21]. Adiabatic circuits inherently avoid such fast switching transitions and therefore reduce this noise source. Consequently, kinetic inductors powering adiabatic circuits (Fig. 1a) based on ultra-steep-slope transistors (Fig. 1b) may represent an advance.

## II. Technical Details

### A. CMOS conversion in detail

Fig. 2 illustrates a process referred to as *CMOS conversion*—defined as upgrading a CMOS design by converting it to reversible logic to reduce energy consumption. Fig. 2a depicts the original CMOS chip as a component within a larger system. Fig. 2b and Fig. 2c illustrate an adiabatic or reversible logic implementation based on the paradigm in Fig. 1a and Fig. 1b, intended to replace the original CMOS component.

In Fig. 2a, the original CMOS chip's internal subsystems (orange and green) are connected via a bus to an external subsystem (blue). The orange subsystem includes an irreversible AND gate. Fig. 2c could retain the same subsystem structure and layout with minor dimensional changes, but the AND gate is replaced by a reversible Toffoli gate.

For the purposes of exposition, consider transforming the original CMOS logic network in Fig. 2a into a partially reversible logic network in Fig. 2c with equivalent functionality, potentially using the techniques in [6]. Reversible gates are placed in positions corresponding to the gates in the original CMOS design. To facilitate comparison, we assume that both designs are fabricated using the same semiconductor process.

This transformation yields two systems with identical functionality but differing logic implementations and power delivery mechanisms (DC versus energy recycling). These systems can be compared in terms of energy consumption and other performance metrics, consistent with the objectives of [4, 13] and related work.

No claim is made that prior work has articulated the full framework in Fig. 2. However, partial implementations may still provide value. Even if the original CMOS is represented only by a block diagram rather than a physical chip, Fig. 2 provides a framework for designing reversible logic systems. Additionally, certain CMOS properties can serve as a planning guide, while allowing design tools to optimize layouts normally.

As an initial step in creating a CMOS-to-reversible comparison, assume that the original CMOS chip datasheet specifies a supply voltage $V_{dd} = V_C$, clock frequency $f_C$, power dissipation $P_C$, and chip area $A_C$, where the subscript C henceforth denotes "CMOS." Using the dynamic power relation $P = ½CV^2 \times f$, an equivalent capacitance $C_C$ can be computed to represent power-related behavior.

The following correction factors are applied (illustrative values shown in parentheses):

- The CMOS conversion process requires dynamic power as an input, but a CMOS data sheet reports total power. The correction factor $\delta_{leak}$ (= 0.8) is the ratio of dynamic power to total power.
- A CMOS data sheet reports total chip area, but the active logic area will be smaller due to "dark silicon." The factor $\delta_{dark}$ (= 2) is the ratio of the total die size to the active die area containing logic.
- When converting a function implemented in CMOS to a reversible gate network of equivalent function, the chip area is likely to increase. The factor $\delta_{size}$ (= 1) captures this proportional increase. This factor has been explored in [6], but will be cited later as an area for further R&D.
- The appropriate power-clock voltage $V_R$ should differ from the original CMOS supply voltage $V_C$ due to CMOS-specific issues like leakage current, use-case constraints such as cryogenic temperatures, and other factors. The resolution of these differences is expressed as $\delta_V$ (= $V_R$ / $V_C$).
- The original CMOS power is spread across four reversible logic power-clocks, assigning $\delta_{qtr}$ (= 1/4) to each phase.

Subsequent sections use CMOS chip parameters to derive reversible logic parameters $A_R$, $P_R$, $C_R$, $V_R$, and $f_R$, where the subscript R henceforth denotes reversible quantities. At present, the reversible logic supply voltage $V_R$ is supplied manually.

In summary, the CMOS conversion approach does not seek to expand the scope of prior ideas [4-5] and recent R&D project, but rather defines more precisely and with a quantitative framework.

### *B. Use of Kinetic Inductors*

The previous section evaluated the energy and power properties of HKI material based on energy density at speed. Here, we describe the patterning of HKI material into kinetic inductors based on the architecture in Fig. 3. Normal metal inductors are often implemented as spirals; however, because kinetic inductors do not rely on magnetic fields, spiral geometries offer no inherent advantage over planar meander structures.

We first consider the central energy-storage region (shown in blue). Let this region comprise wires of width $w$ separated by a minimum wire-to-wire spacing $s_{min}$. In power-handling applications, $w$ typically exceeds the minimum feature size $w_{min}$. Neglecting integer effects, the energy storage region can be modeled as a series connection of repeating units, each consisting of a $w \times w$ square of HKI material and an adjacent $w \times s_{min}$ insulating region. The resulting inductance density is $L_\square/(w \times (w + s_{min}))$ H/m$^2$. Although the inductance of the energy-storage region varies with $w$, the energy capacity is primarily affected by the reduction in active HKI material due to spacing gaps. This effect becomes asymptotically smaller as $w$ increases.

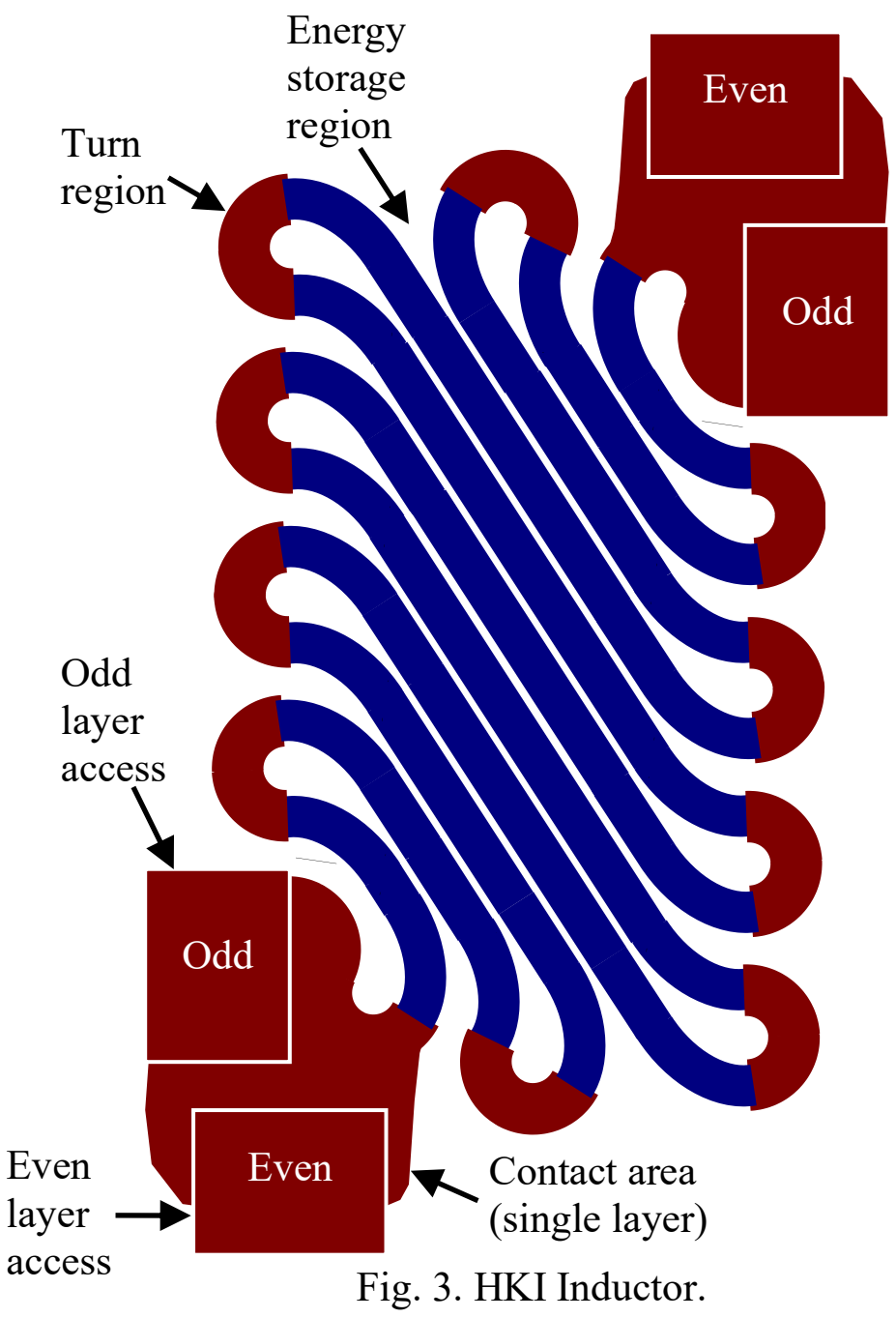

Fig. 3. HKI Inductor.

The semicircular features shown in red redirect the current. Charge carriers propagate in straight paths unless redirected by boundaries, leading to current crowding that depends on turn sharpness. To mitigate this effect, the architecture features widening conductor widths to reduce the nominal current below $I_c$, and turns designed with a minimum radius so that the peak current remains below the critical current even under crowding conditions. These considerations are illustrated in Fig. 3. Kinetic inductors for other applications (e.g., qubit resonators) are typically optimized with 2D simulators, which would apply to this architecture as well.

This analysis becomes increasingly accurate under asymptotic conditions. As the total inductor area increases, the border area scales linearly, while the energy-storage area grows quadratically, making the latter dominant. Furthermore, as the ratio $w/s_{min}$ increases, the impact of wire spacing gaps diminishes. These observations justify the use of an area-based approximation for HKI inductor energy capacity.

### *C. Quantification of CMOS Conversion*

This section describes CMOS conversion structured around a sequence of equations utilizing representative numerical values. The narrative first sets up the problem using a set of equations and then partitions the energy-recycling power supply into regions. These regions are analyzed for different inductor types and use cases. Numerical values are introduced using parenthetical expressions beginning with an equal sign, e.g., $P_i$ (= 3.1415).

The equations are subsequently reapplied with alternative parameters to assess alternative scenarios. A spreadsheet and Python script are included in the supplemental materials.

Consider an original CMOS design with area $A_C$ (= 1 cm$^2$), power supply voltage $V_C = V_{dd}$ (= 1 V), clock frequency $f_C$ (= 1 GHz), and power dissipation $P_C$ (= 100 W), along with the additional properties defined below.

Using the standard CMOS dynamic power relation ($P = ½CV^2 \times f$), the equivalent capacitance is:

$$C_C = 2P_C / (V_C^2 f_C) \text{ } (= 0.2 \text{ µF}),$$

where $C_C$ denotes equivalent capacitance of the CMOS circuit for power analysis.

Applying the correction factors defined above yields

$$A_{qtr} = A_C \times \delta_{qtr} \text{ } (= 0.25 \text{ cm}^2),$$

$$C_{qtr} = \delta_{leak} \times \delta_{dark} \times \delta_{size} \times \delta_{qtr} \times C_C \text{ } (= 0.08 \text{ µF}),$$

where the subscript "qtr" henceforth denotes quantities associated with one of the four regions (inductors) in Fig. 2b. Here, $A_{qtr}$ is the reversible logic area allocated per phase and $C_{qtr}$ is the corresponding CMOS capacitance after corrections.

*1) Integrated Inductors and Scaling to a Target Frequency*

By the definition of integration, each inductor in Fig. 2 is the same size as the reversible logic circuitry beneath it. The resulting resonant frequency is typically too low. The remedy presented in this document is to subdivide the region into $N_R$ equal-sized subregions, as illustrated in Fig. 4 for $N_R$ = 12. Because capacitance is assumed to scale linearly with area and inductance increases with area (for all inductor models used here), subdivision increases the resonant frequency.

This section derives expressions for $N_R$ based on different inductor types. The first step computes an initial inductance value and a scaling exponent.

*2) Kinetic Inductors at Minimum Conductor Width $w_{min}$*

Representative parameters for low-$T_c$ superconducting kinetic inductors [15] are $L_□$ (= 8.5 pH/□), $w_{min}$ (= 0.8 µm) minimum wire width, $s_{min}$ (= 1 µm) minimum wire spacing. The initial inductance is:

$$L_k = L_□ \times A_{qtr} / (w_{min} \times (w_{min}+s_{min})) \text{ } (= 148 \text{ µH}),$$

where the subscript k henceforth denotes "kinetic." Since $L_k$ scales linearly with area, the scaling exponent is $\Sigma_k = 1$.

*3) Meander Inductors*

For normal metal and non-kinetic superconductors, an appropriate monomial expression for meander inductance [22] is:

$$L_m = 0.0026\, a^{0.0603}\, h^{0.4429}\, N^{0.954}\, d^{0.606}\, W^{-0.173},$$

where the subscript m henceforth denotes "meander," $a$ is lead length, $h$ is meander height, $N$ is number of meanders, $d$ is meander spacing, $W$ is wire width, and $s = 0$ ($s$ is a parameter in [22]).

In this document, an asymptotically dense square geometry is defined with $a = d/2$, $l = len = h + W$, and $W = d$, yielding

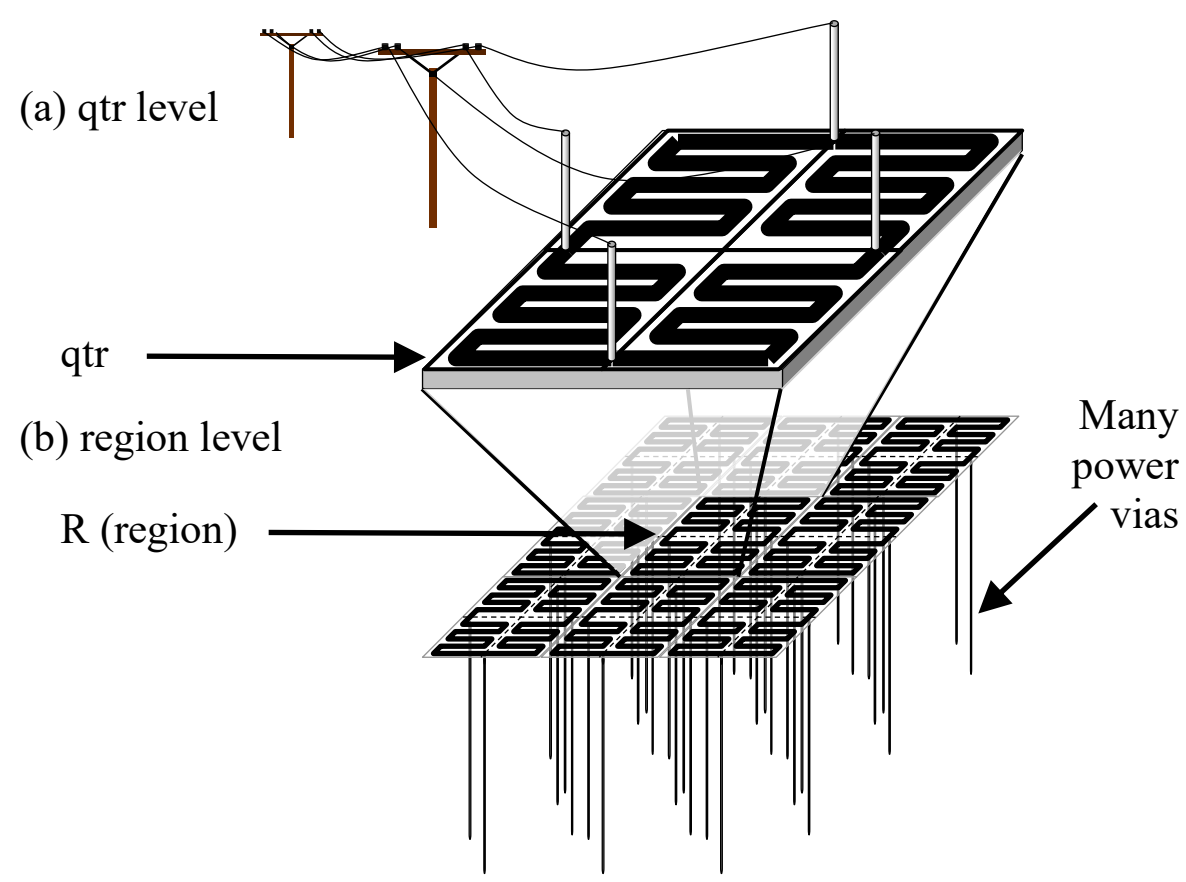

Fig. 4. Regions and tuning to a specific frequency

$$L_m \text{ } (= 144 \text{ nH}).$$

For a fixed $N$, $L_m$ scales as $A^{0.4681}$, hence $\Sigma_m = 0.4681$.

*4) Spiral Inductors*

For normal metal and non-kinetic superconductors, an appropriate expression for spiral inductance [23] is:

$$L_s = 0.635\, \mu_0 N^2 d_{avg}[\ln(2.07/\rho) + 0.18\rho + 0.13\rho^2],$$

where subscript s henceforth denotes "spiral." In this document, an asymptotically dense geometry with $d_{avg} = 0.635$ and $\rho = 1$ ($d_{avg}$ and $\rho$ are parameters in [23]), yields

$$L_s \text{ } (= 63 \text{ nH}).$$

For fixed $N$, $L_s$ scales as $A^{½}$, hence $\Sigma_s = 1/2$.

*5) Scaling Area and Frequency*

The resonant frequency of $L_{type}$ and $C_{qtr}$ generally differs from the target frequency $f_R$. To reach $f_R$, $A_{qtr}$ is subdivided into smaller regions. As the region area decreases, both capacitance and inductance decrease, thereby increasing the resonant frequency. The initial resonant frequency is

$$f_{qtr_type} = \sqrt{2} / (2\pi \sqrt{L_{type} C_{qtr}}),$$

where "type" in a subscript henceforth denotes the inductor type (k, m, or s), and $f_{qtr_type}$ is the resonant frequency when the area of the inductor and capacitor is $A_{qtr}$. The $\sqrt{2}$ arises from the 4LC quadrature mode, which is detailed extensively below.

The key scaling step (Fig. 4) is the subdivision of qtr into regions. This step scales the inductors by area and connects them to the reversible logic circuitry underneath, replicating the subdivision to fill the original area. Since the reversible logic circuit is presumed to have uniform capacitance per unit area, this step reduces capacitance per region by the replication factor. The required number of regions is:

$$N_{Rtype} = (f_R / f_{qtr_type})^{2/(1+\Sigma type)},$$

where $N_{Rtype}$ is the number of regions needed to raise the resonant frequency to $f_R$, for each inductor type. Non-integer values of $N_{Rtype}$ imply an approximation.

Using the parameters and equations above:

$$N_{Rk} \text{ } (= 1{,}527) \text{ (for kinetic inductors)},$$

$$N_{\mathrm{Rm}} \ (= 40) \text{ (for meander inductors),}$$

$$N_{\mathrm{Rs}} \ (= 100) \text{ (for spiral inductors),}$$

assuming $N = 4$ meanders or turns.

*6) Non-Superconducting Meanders and Spirals*

To prevent artificially inflating performance metrics within the planning model, this document defines:

$$h_{\mathrm{tlm}} = \min(10\ \mu\mathrm{m}, w),$$

where $h_{\mathrm{tlm}}$ is metal thickness and $w$ is wire width.

For calculating inductor resistance and thus the Q factor, the model uses the wire width that establishes the correct resonant frequency, but thickness is capped at 10 µm. This prevents the model from predicting unrealistic performance by specifying an unmanufacturably thick metal layer.

For meander inductors:

$$L_{\mathrm{Rm}} \ (= 2.55 \text{ nH}),$$

where $L_{\mathrm{Rm}}$ is the inductance computed from the meander formula applied to area $A_{\mathrm{qtr}} / N_{\mathrm{Rm}}$,

$$C_{\mathrm{Rm}} = C_{\mathrm{qtr}} / N_{\mathrm{Rm}} \ (= 1.99 \text{ nF}),$$

$$R_{\mathrm{Rm}} = (2\,N\,h + 2\,h) / (h_{\mathrm{tlm}}\,w) \times \rho_{\mathrm{copper}} \ (= 0.189\ \Omega),$$

where $R_{\mathrm{Rm}}$ is the resistance of the meander, calculated from the wire length ($2\,N\,h + 2\,h$) divided by the cross-sectional area ($h_{\mathrm{tlm}}\,w$), multiplied by copper resistivity $\rho_{\mathrm{copper}}$ ($=1.68\times10^{-8}$).

For spiral inductors:

$$L_{\mathrm{Rs}} \ (= 6.33 \text{ nH}),$$

where $L_{\mathrm{Rs}}$ is the inductance computed from the spiral formula applied to area $A_{\mathrm{qtr}} / N_{\mathrm{Rs}}$,

$$C_{\mathrm{Rs}} = C_{\mathrm{qtr}} / N_{\mathrm{Rs}} \ (= 0.8 \text{ nF}),$$

$$R_{\mathrm{Rs}} = \tfrac{1}{2} \sqrt{(A_{\mathrm{qtr}} / N_{\mathrm{Rs}})} / (h_{\mathrm{tlm}}\,w) \times \rho_{\mathrm{copper}} \ (= 0.214\ \Omega),$$

where $R_{\mathrm{Rs}}$ is the resistance of the spiral, derived from half the edge length times four times the number of turns divided by cross-sectional area ($h_{\mathrm{tlm}}\,w$) multiplied by copper resistivity.

Based on the quality factor equation $Q = 1/R \times \sqrt{(L/C)}$:

$$Q_{\mathrm{Rm}} \ (= 5.98) \text{ (for meanders),}$$

$$Q_{\mathrm{Rs}} \ (= 13.15) \text{ (for spirals).}$$

These values help explain the persistence of an "infinite loop" of projects attempting to realize reversible logic test chips. Under idealized assumptions—specifically, that inductors are the sole source of loss—the maximum energy efficiency improvements are 5.98× and 13.15×. A commonly cited "10× rule" suggests that displacing an entrenched competitor requires at least a 10-fold improvement to motivate customers to overcome switching costs. Only one of the computed values exceeds this threshold, and additional loss mechanisms have not yet been included.

*7) Superconducting but Non-Kinetic Meanders and Spirals*

Meander and spiral inductors fabricated from superconducting wire have zero resistance; however, superconductivity breaks down when current exceeds the critical current $I_c$. In this document, $I_{\max} = I_c \times w$, where $I_{\max}$ is the maximum current in the wire, $w$ is the width of the wire, and $I_c$ is defined per unit width for a wire of a specified height.

Representative performance of a spiral YBCO inductor [19], as used in an ion trap quantum computer, features a maximum current of 0.64 A through a 300 µm-wide wire, corresponding to $I_{c,\mathrm{traphub}}$ = 2,133.33 A/m. Based on the parameters and equations presented above, the corresponding utilization is 75% for meander inductors and 48% for spiral inductors. These high-$T_c$ inductors exhibit electrical characteristics consistent with [5], featuring $Q_{\mathrm{traphub}}$ = 1,172, expanding the efficiency budget available to other loss mechanisms within a reversible system.

One option for future work would be to use the tested and characterized high-$T_c$ inductor design or existing fabricated artifacts (if available) [19].

Alternatively, discussions with the foundry that created [19] indicate they accept engineering designs for fabrication with similar materials. A YBCO chip could potentially be fabricated with four 4LC circuits to physically test the resonator presented in this document.

### *D. Cryo-CMOS Conversion*

Assuming at least a few HKI layers can be fabricated [15], Table 1 demonstrates that one or two layers are sufficient to perform CMOS conversion on cryo-CMOS quantum-computer controllers [16, 17]. Table 1 uses the equations above but override specific parameters as shown below, yielding the required number of HKI layers.

Table 1. CMOS conversion of a representative [16] qubit controller to reversible logic

| | | |
|---|---|---|
| $f_R$ = 100 MHz | $f_R$ = 100 MHz | Operating frequency (qubits only) |
| $V_R$ = 0.4 V | $V_R$ = 0.497 V | $V_R$ based on $V_C$ = 1 V |
| $P_C$ = 13 W | $P_C$ = 17 W | $P_C$ based on original 100 W |
| $L_\square$ = 11 pH/□ | $L_\square$ = 11 pH/□ | Inductance per square, was $L_\square$ = 8.5 pH/□ |
| $w$ = 28.8 µm | $w$ = 34.4 µm | Wire width, was 0.8 µm |
| 1 layer | 2 layers | Computed number of HKI inductor layers |

Thus, it should be possible to convert a cryo-CMOS controller [16, 17] to four-phase reversible logic, subject to the following considerations:

- The conversion applies only to the CMOS logic [16], not the analog circuits.
- The conversion applies only to the operating speed required for interfacing with qubits (100 MHz) [16]. While testing of the cryo-CMOS chips [16] may utilize speeds of 1 GHz or more, these higher speeds are not strictly necessary.

Under the assumptions of the planning framework developed above, these results suggest that a reversible redesign may achieve advantages over existing cryo-CMOS designs in metrics such as power dissipation and switching-noise generation. Consequently, cryogenic quantum-control electronics may represent one of the first application domains in which reversible logic could become commercially relevant.

The ~4 K cryo-CMOS controllers interface with qubits operating at millikelvin temperatures where low power dissipation is critical. While the following point remains to be demonstrated in context, qubits are extremely noise-sensitive and harsh CMOS transitions are cited in the literature as a source of this noise; the power-clocks used in reversible logic simply do not have the corresponding transitions and are therefore expected to yield lower switching-noise generation.

An alternative implementation is to employ ultra-steep-slope CMOS transistors in the reversible redesign. Kinetic inductance remains stable at temperatures down to approximately 0.5 K [17], where a substantial reduction in $V_{dd}$ would decrease the required number of HKI layers, or alternatively support scaling to larger qubit arrays.

Without reversible logic, CMOS—including ultra-steep-slope CMOS [17]—produces multi-GHz switching noise that can interfere with sensitive signals [21], including qubits and sensor signals. Several use cases illustrate the potential advantages:

(a) Quantum computer control: Increased qubit count and local control. Local control improves precision by operating at the same temperature stage, thereby enhancing qubit fidelity.

(b) Sensor arrays with integrated readout: Lower operating temperatures enable *in situ* operation directly alongside sensor elements, supporting progressively longer-wavelength imaging.

At the time of this writing, quantum-computer development is a national priority. For example, DARPA's Quantum Benchmarking Initiative (QBI) aims to achieve a "utility-scale" quantum computer by 2033 [24]. The concepts presented in this document may provide a path toward more energy-efficient control electronics, potentially enabling larger qubit counts within the same controller power budget.

### E. Tailoring the Energy-Recycling Power Supply

Having established that integrated kinetic inductors provide sufficient energy capacity, the circuit in Fig. 1a is now tailored to the CMOS conversion process. Lithographically defined kinetic inductors can be subdivided into smaller elements without reducing energy density per unit area, provided the asymptotic assumptions described above remain valid.

We first consider the quadrature circuit shown in Fig. 5, denoted as the 4LC circuit. This circuit can be interpreted as a four-stage lumped-element transmission line arranged in a cycle, although it is analyzed here as a general LC network. One resonant mode of the 4LC circuit consists of sinusoidal waveforms in quadrature, which can serve as the four-phase ramped power-clocks used in T-gate logic [8], a.k.a. 2LAL [9]. Prior work has explored the use of sinusoidal waveforms in place of ramped waveforms for reversible logic power-clocks; simulations conducted for this document, included in the supplemental material, indicate comparable energy efficiency.

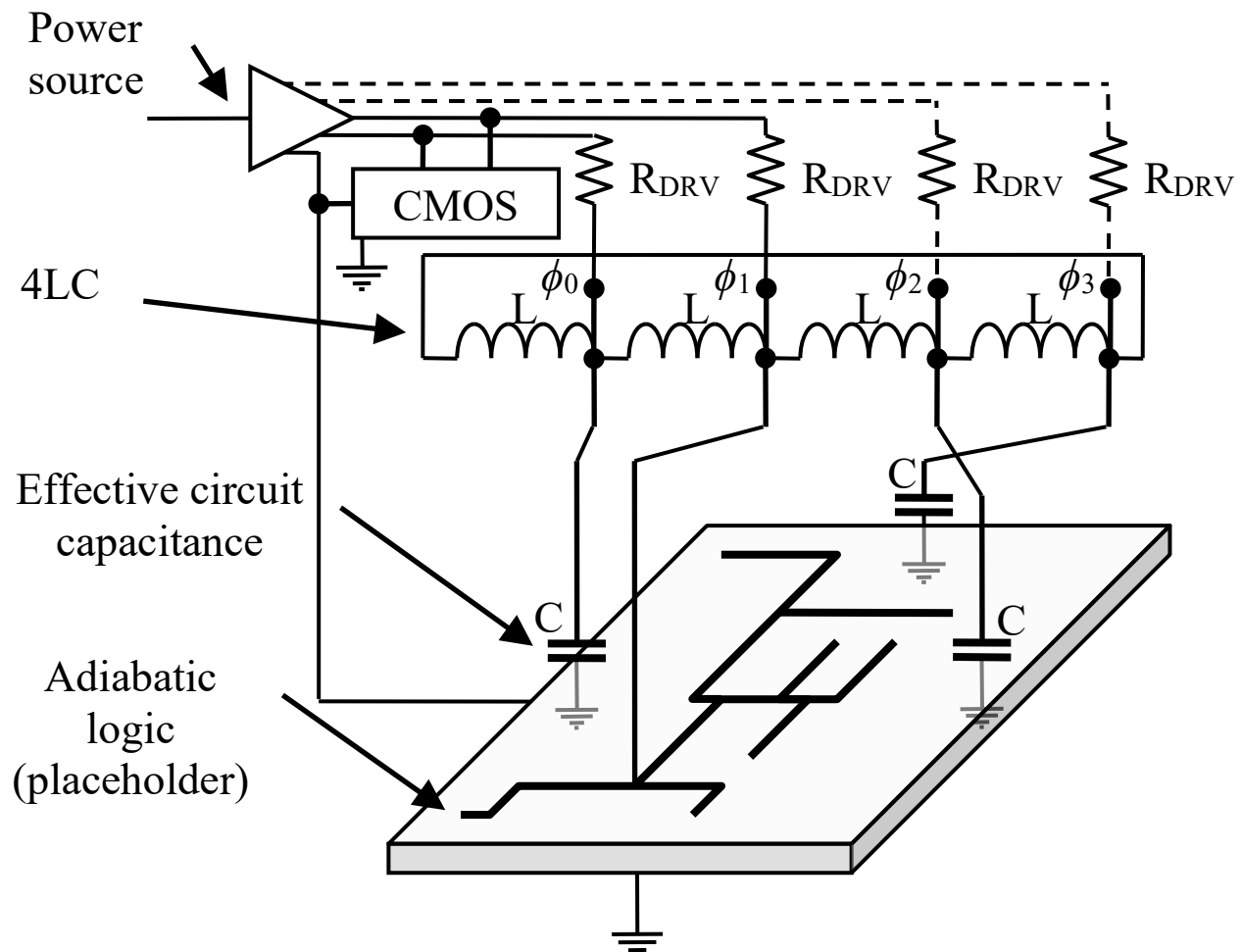

Fig. 5. Energy recycling system diagram.

The 4LC circuit must exchange energy with an external power source, which may employ conventional electronics to generate drive waveforms matching the desired power-clocks. These waveforms are injected through resistors ($R_{DRV}$ or generally $R_{DRV0}$–$R_{DRV3}$), as shown in Fig. 5. The resistors enable controlled energy transfer into and out of each oscillatory mode until the amplitude and phase align with the externally driven waveforms.

Appendix 3 describes the transient simulation of the energy-recycling power supply in Fig. 5, analyzing how the 4LC circuit drives an underlying reversible logic circuit.

### F. Chip Power Grid

The preceding discussion demonstrated that four inductors can drive four-phase universal logic (e.g., T-gate or 2LAL) without active semiconductor switching in the power-clock generator. Since the inductors in this document are created lithographically, more than four inductors can be created, providing further architectural advantages.

As illustrated in Fig. 2b, consider converting a 100 W chip by connecting the four power-clocks to the midpoints of the four edges of the die.

Each phase may carry currents on the order of tens of amperes. Distributing this current across the chip requires a carefully designed power distribution network to minimize losses. Positioning the energy storage elements within a few micrometers of the points of energy usage significantly simplifies this network.

To achieve this, the chip may be subdivided into $N$ regions, with $N = 4$ used here for illustration.

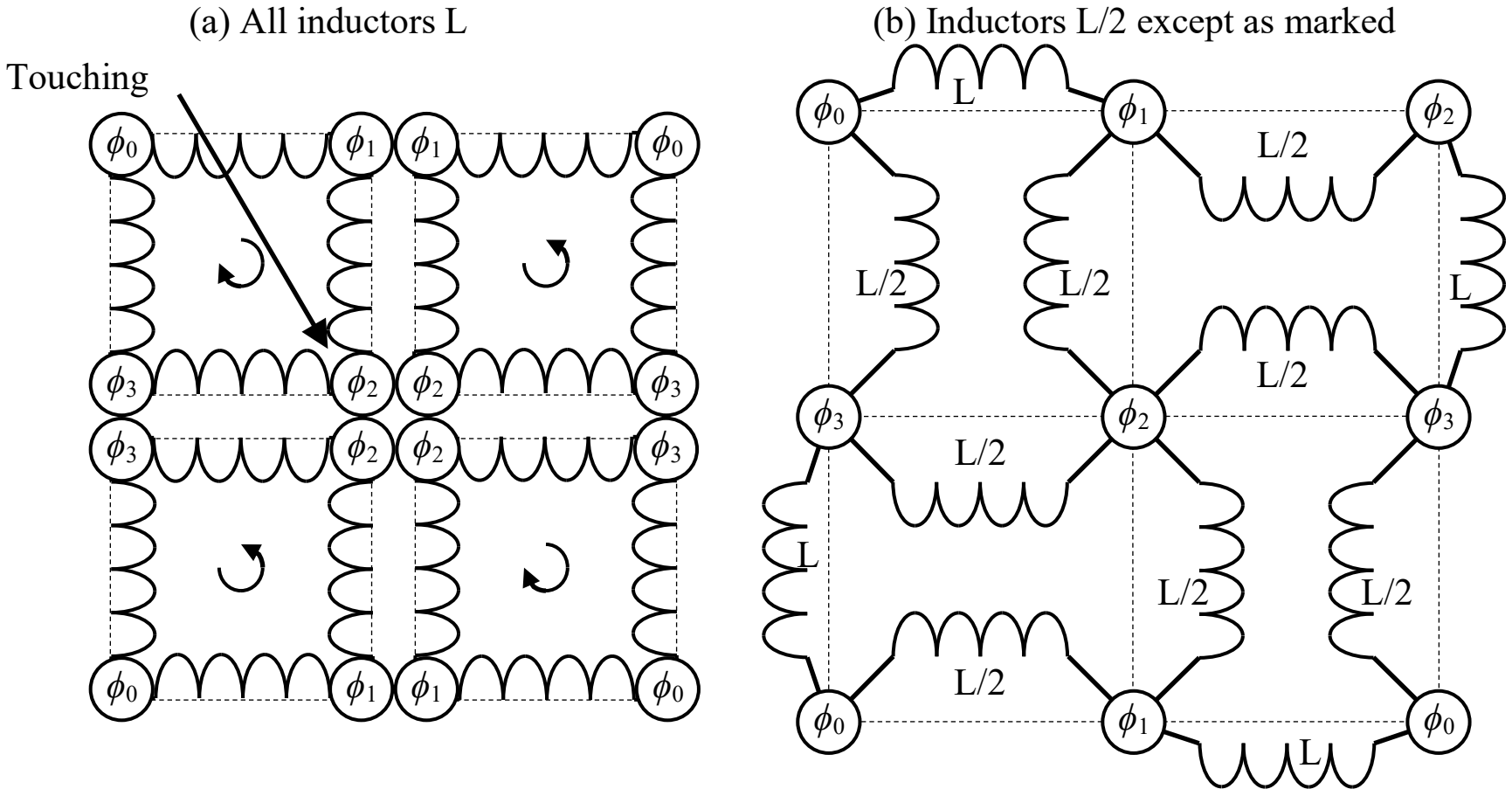

Fig. 6. Chip power grid. (a) Think about placing 4LC circuits next to each other. (b) Combine parallel inductors into two types of tiles with inductors of L/2. Inductors of L also needed on the boundary.

The baseline approach implied in Fig. 4 for the parameter $N_R$ assigns an identical 4LC circuit to each region; this section can be considered a second application of the same subdivision principle.

As shown in Fig. 6a, arranging the 4LC circuits in a checkerboard pattern by rotating and mirroring them results in adjacent power-clock nodes (illustrated as touching circles) sharing the exact same phase. If all 4LC circuits are initialized and driven identically, their nodal waveforms will be identical. Consequently, nodes with the same waveforms can be merged into a single electrical node without affecting macroscopic circuit operation.

Fig. 6b illustrates a refinement of this power grid. Adjacent inductors of value $L$ are merged into equivalent shared inductors of value $L/2$, while the edge inductors remain at $L$. The merged inductors are reassigned to neighboring 4LC circuits, with vertical inductors associated with clockwise numbering and horizontal inductors with counterclockwise ordering.

The resulting power grid forms a checkerboard pattern comprising two distinct subcircuits: one containing vertical inductors with clockwise phase numbering, and the other containing horizontal inductors with counterclockwise phase numbering.

Superconducting wire forms both the kinetic inductors and the interconnect leads connecting them. This approach delivers power over superconductors directly to points adjacent to the relatively lossy semiconductors, substantially reducing power distribution overhead.

The proposed chip power grid does not require uniform region shapes, equal areas, or identical capacitances. It is applicable to arbitrary physical partitions, provided that sufficient routing resources exist to implement a properly tuned 4LC circuit for each region at the desired operating frequency.

In practical designs, regions may exhibit highly non-uniform power density (e.g., high-power arithmetic units versus lower-power memory blocks). In such cases, inductors supplying high-power regions may extend physically into adjacent lower-power regions to improve layout efficiency.

Regular grid structures benefit from merging contact points and inductors, thereby maximizing layout density. Conversely, highly nonuniform workloads may favor irregular partitioning, even if minor additional routing is required to maintain system connectivity.

In summary, the chip power grid approach assigns an independent 4LC circuit to each localized region, based on its specific capacitance and the inductance required for resonance. Layout is then optimized by merging power-clock nodes, consolidating inductors between common nodes, and place-and-route of inductor structures directly above their corresponding logic regions.

## III. Designing with the 4LC Circuit

This section discusses design methodology for the 4LC circuit. Assuming the designer has utilized the planning equations and is assured that the kinetic inductors possess sufficient energy capacity to power the logic, the primary task is to understand the load that a reversible logic circuit places on the 4LC power supply. This includes the load of the idealized circuit, manufacturing variances, and the effect of data in the computation. This approach determines specific component values subject to the statistical model of the circuit demonstrating how to simulate performance across statistical sweeps.

This section develops co-design principles for the 4LC and reversible logic circuits based on the capacitance $C_{qtr}$ of reversible logic, previously introduced in the equation $C_{qtr} = \delta_{leak} \times \delta_{dark} \times \delta_{size} \times \delta_{qtr} \times C_C$, where $C_C$ is the circuit capacitance of the original CMOS circuit. However, the reader will recall that the variable $\delta_{size}$ has not yet been assigned a specific value. This section will define $C_{qtr}$ based on an RLC circuit analysis, providing a method for determining $\delta_{size}$ and the method for computing it from the analysis of a reversible logic family, such as the one used in Fig. 1b.

Modeling $C_{qtr}$ exclusively from each power-clock to ground served as an initial convenience. The 4LC power supply actually connects to the logic circuit through four power-clock wires plus GND. There are ten possible pairs among these five wires, each of which could have an associated wire-to-wire capacitance. However, due to circuit symmetry, there are only three independent values for these capacitances: $C_{GND}$ for a phase to GND, $C_{90}$ between adjacent phases, and $C_{180}$ between opposite phases. While these new capacitors may seem to replace the 4LC with a more complex circuit, Appendix 4 demonstrates that the circuit's behavior depends only on a weighted sum of these capacitors, which we now define as $C_{qtr}$.

The equivalent logic circuit model is inherently dependent on the processed data. Wires and transistor gates are the sources of a reversible logic circuit's capacitance, but these capacitances are driven by different clock phases over time based on the data being processed. Each wire or gate may be driven by two power-clocks of opposite phases (e.g., phases 0 and 2). The connection to the driving power-clocks is made through transmission gates controlled by the other phases (e.g., phases 1 and 3). Unless floating, a wire's capacitance contributes to the load of one of these driving power-clocks.

Consequently, the switching of capacitance in and out of a power-clock's load translates the processed data into "capacitance noise."

The designer can partially control capacitance noise at the circuit level. For example, a dual-rail signal generates both true and complement signals, such as A and –A. Both rails tend to connect the same driving and receiving gates, allowing the designer to use similar geometric shapes for the two wires. Given that no more than one wire is connected at any time, the capacitance noise of the dual-rail connection will be $\Delta C = |C_A - C_{-A}|$, which the designer can minimize through careful layout or use of a suitably programmed routing tool. However, reversible busses and data-controlled reversible power-clocks, introduce capacitance noise that cannot be mitigated as easily.

We assume a design tool will compute the amount of capacitance for each wire and divide $C_{qtr}$ into statistically varying capacitive loads ($C_{GND}$, $C_{90}$, and $C_{180}$). These capacitance values can be used to simulate the 4LC circuit—including simulations over many samples of capacitive noise. If the simulation reveals excessive noise, reducing the drive resistors $R_{DRV}$ will damp transients more quickly—albeit at the expense of requiring the AC power source to have lower output impedance.

Appendix 4 details this analysis. This process can be applied to a specific circuit, yielding a value for $\delta_{size}$ alongside other relevant details, such as the ratio between $C_{90}$ and $C_{180}$. Furthermore, the process could be applied to a broader design style in conjunction with a specific semiconductor process. The value of $\delta_{size}$ is essentially a statistical design efficiency metric resulting from the wiring layers and parameters of a semiconductor process and the circuit complexity of the reversible logic family being used—so it should be about the same for a microprocessor, GPU, or other digital circuit. Knowing the value of $\delta_{size}$ would allow for feasibility assessments and could serve as a standard metric. While generally beyond the scope of this document, the analysis in Appendix 4 could provide insight into how suitable a semiconductor process might be for reversible logic, perhaps establishing a roadmap for improving the process.

Appendix 4 also models kinetic inductor self-resonance as a capacitor across each inductor's terminals—which becomes an additional component in $\delta_{size}$ with the same implications outlined above.

Additionally, Appendix 4 develops a transient simulation method, complete with an example, that computes the effects of data variance on the efficiency of the 4LC circuit. In summary, the simulation method performs an RLC simulation for one phase at a time (i.e., a quarter of a clock period). The simulator modifies capacitance values between ticks to model data-dependent load changes. This does not require a nodal analysis between ticks; rather, it requires recreating the 8 × 8 system matrix using expressions provided in the appendix.

Fig. 7 illustrates the final result from Appendix 4. The top two colored bands represent losses in the reversible logic, specifically from opposite and adjacent phases. The smaller colored bands at the bottom represent losses in the drive resistors, which damp out spurious modes caused by data-induced load changes.

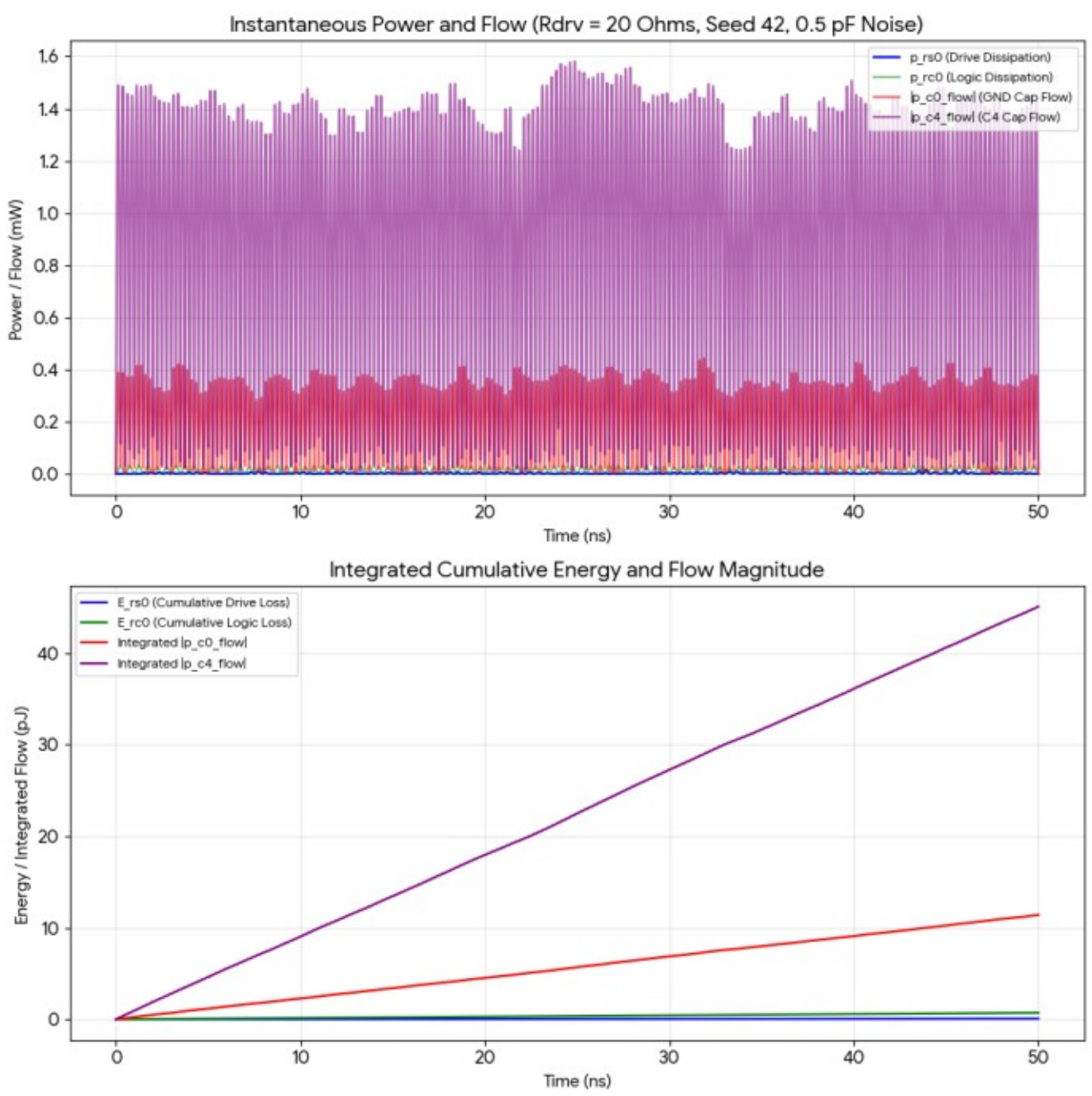

Fig. 7. Energy recycling including capacitance noise.

The bottom chart shows growing energy consumption over time, revealing the effectiveness of energy recycling. The top line represents the energy entering the reversible circuit, which is roughly the counterpart of CMOS energy consumption. The second curve represents energy dissipated in the damping resistors $R_{DRV}$. Whereas CMOS turns all input energy into heat, reversible logic recycles it. The lower two curves represent the dissipation in both the reversible logic and the damping resistors, which together account for roughly 1/30 of the top curve—indicating 29/30th of the energy is recycled from one clock to the next.

## IV. Conclusions and Future Work

Reversible logic projects have produced about a dozen test chips that were not followed by commercially viable systems. To make progress, this document described a general top-down framework for CMOS conversion to reversible logic. The planning formulas were applied to normal metal inductors (Fig. 1a), revealing that they consume all or nearly all of the loss budget, leaving little room for other necessary functions. This justifies improved inductors as a valid future direction.

The document introduced kinetic inductors (Fig. 1b), which provide substantially greater energy density and Q factor and appear capable of removing a major scaling constraint identified by the framework. The framework subdivides the scale-up process into a series of quantifiable issues, each one of which should be considered before funding future projects, and would reduce upfront risk and provide mid-term tests for development projects.

Enabling scale-up does not imply that reversible logic will beat CMOS for every application. This document identified some current cryo-CMOS quantum computing controller chips as amenable to conversion to reversible logic using available technology, with a strong prospect of outperforming CMOS on relevant metrics (less heat generated in the cryostat and lower noise). This document does not conclude that reversible logic will be a universal replacement for CMOS, but rather that the planning formulas reveal that other categories of applications with similar computational and environmental requirements could become viable as technology improves.

The framework includes a 4LC resonator (Fig. 5) capable of generating four-phase power-clocks directly, a distributed power-distribution mesh (Fig. 6) that reduces loss through very short power wires (Fig. 4), and a synthesis algorithm that outputs a circuit (Fig. A-7) capable of converting the four-phase power-clocks into potential power-clocks with more phases or retractile cascades.

The document disclosed RLC analysis of the 4LC power supply (replacing Fig. 1a) in conjunction with an adiabatic logic circuit (Fig. 1b and alternatives [5-13]).

The tight coupling between kinetic inductors and the reversible logic circuit removes some opportunities for power loss—but introduces a low-loss path for noise generated by the reversible logic to enter the resonator. Prior work has examined load effects in adiabatic and resonant power-clock systems; however, the authors are not aware of prior work that co-designs the resonator with capacitance noise from data in the computation to optimize energy efficiency.

### *A. Future Work*

While this document demonstrated how kinetic inductors enable reversible logic, it did not include fabricating a chip. Reaching commercial viability will require a scale-up sequence. Each such chip could be based on a microprocessor or FPGA block [6] using known reversible logic circuit families [8–13] or a newly developed family, including those based on retractile cascades. Initially, 4 K kinetic inductors could be used. The methods in this document describe the kinetic inductor, which could be fabricated by facilities such as SeeQC [15] or MIT-LL. The values of $C_{90}$ and $C_{180}$ would be extracted from the layout geometry and simulated. To continue the scale-up sequence, the same circuit family would be used for circuits of growing complexity. As the process proceeds, statistical values for parameters (e.g., $C_{90}$ and $C_{180}$) would become better characterized the improved values could be incorporated into design tools.

The impact of the $\delta_{size}$ correction factor underscores a critical need for more efficient reversible logic families. Prior work [6] analyzed the scaling impact of $\delta_{size}$ across both "asynchronous" logic families and retractile cascades, demonstrating that retractile cascades yield superior area efficiency. Consequently, future research could explore a heterogeneous system architecture where the 4LC power-clocks are adapted to simultaneously drive a mixture of retractile cascades and asynchronous logic families. This approach would allow individual subfunctions to be implemented using the move efficient logic family for a given workload, provided that data can be seamlessly exchanged across the differing logic domains.

Future work includes detailed circuit validation and multilayer integration strategies, including integration of the superconductor devices described in this document with a semiconductor (CMOS) process. This would initially use a 4 K process such as the one in [15], followed by the exploration of high-$T_c$ processes, such as those from STAR Cryoelectronics [19]. The CMOS process could use ultra-steep-slope transistors, such as from SemiQon [20], and could be extended to include 3D kinetic inductors.

## APPENDIX 1: GENERAL CLOCK GENERATION

Up to this point, this document has described a four-phase clocking scheme. We now extend this approach to an arbitrary number of phases, including retractile cascades [6]. Fig. A1-1 illustrates a four-to-eight-phase converter; however, the method generalizes to four or more phases.

The proposed method defines an energy-recycling four-phase waveform alphabet (Fig. A1-1c) consisting of DC power, GND, an open circuit (tri-state) condition, and power-clock waveforms. These power-clocks may be implemented as conventional ramped waveforms, sinusoidal waveforms, or waveforms with other transition profiles.

The method employs a set of 4*N* dual-rail phase selectors, including active-high selectors (Fig. A1-1a) and complementary active-low versions (not shown). Each selector is active during one of the 4*N* phases.

The circuit is constructed using bidirectional transmission-gate selectors, represented as single transistors in Fig. A1-1b for simplicity. Each selector connects one waveform from the alphabet to the output wire during its assigned phase interval. Because both single transistors and transmission gates are bidirectional, the resulting output waveform supports energy recycling. The configuration shown in Fig. A1-1 produces an eight-phase clock of the type used for S2LAL [10].

This framework may be interpreted as a waveform language in which an *M*-tick waveform is represented by a sequence of *M* symbols. The choice of symbols is arbitrary; however, the mapping in Table 2 generates the eight-tick waveform (in Fig. A1-1b) for the eight-symbol word 51545359.

The formal grammar and semantics of this language are beyond the scope of this document.

Table 2: Waveform to Symbol

| Connect output wire to … | Symbol |
|---|---|
| GND | 9 |
| $\phi_0$ | 0 |
| $\phi_1$ | 1 |
| $\phi_2$ | 2 |
| $\phi_3$ | 3 |
| $V_R$ | 4 |
| open circuit | 5 |

A generator for arbitrary power-clocks with four or more ticks may be synthesized as follows: Given a target waveform $W_1$, construct extended waveforms $W_2$ and $W_4$ by concatenating two and four copies of $W_1$ respectively. Select the shortest waveform whose length is a multiple of four (i.e., 4 *k*) and denote it as *W*. Extend the circuit of Fig. A1-1 to 4 *k* ticks and use it to synthesize a waveform generator for *W*.

This process supports the generation of retractile cascades, data signals, bus driving waveforms (including tri-state behavior), and input/output signals (e.g., LED drivers).

## APPENDIX 2: MULTILAYER KINETIC INDUCTORS

Three-dimensional flash memory is fabricated using a process such as stair-step etching [18]. This process first creates a homogeneous (non-patterned) multilayer structure through alternating deposition of materials (e.g., insulator and semiconductor). A single lithographic step defines features across all layers simultaneously, followed by a stair-step etch that enables electrical access to individual layers. This approach has reached hundreds of memory layers, increasing the key performance metric for memory (i.e., density) by the same factor. The present opportunity is to apply the same approach to the energy capacity of inductors (see [14]).

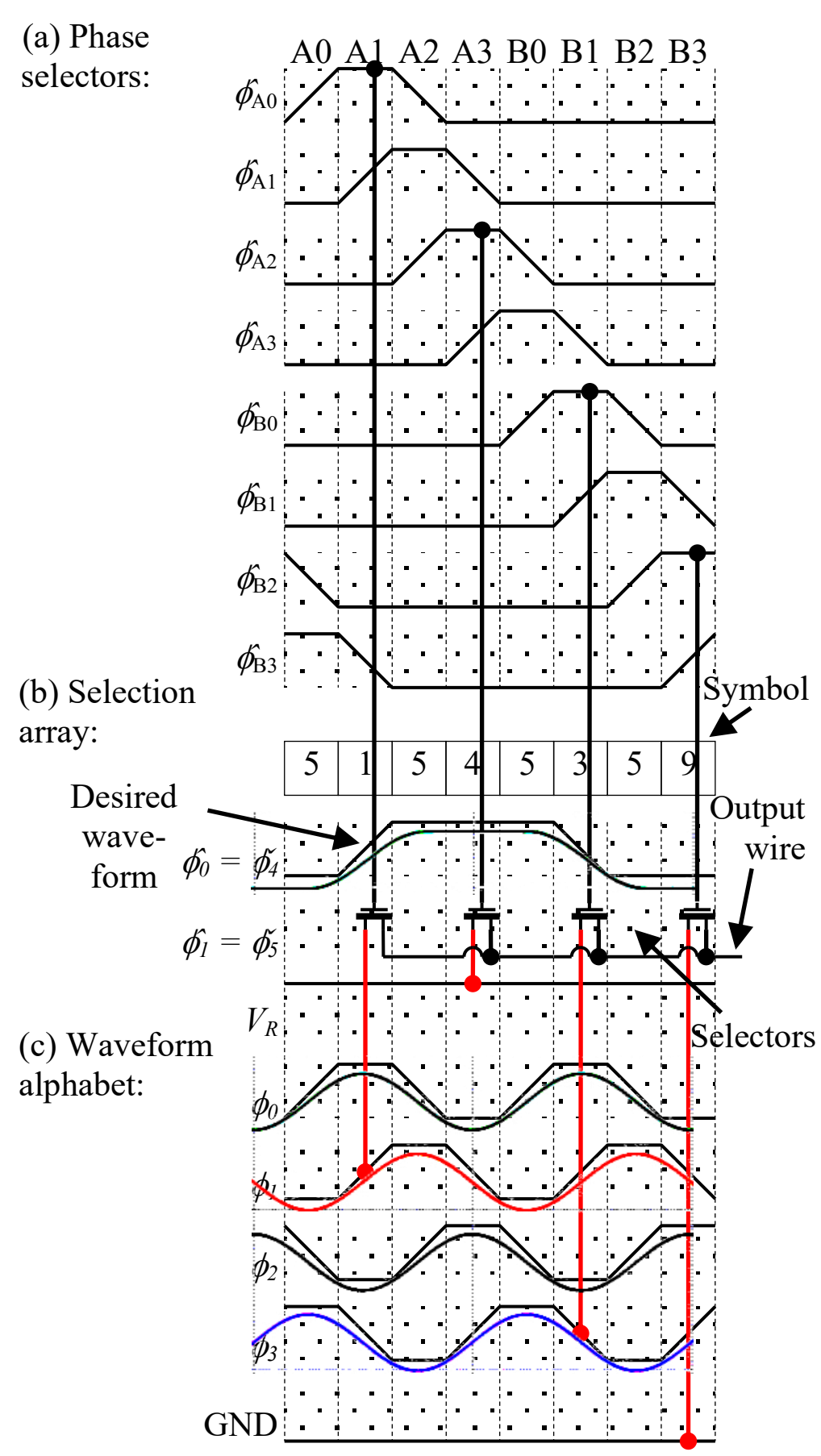

Fig. A1-1. Four-to-eight-phase converter.

The first aspect of the multilayer kinetic inductor concept is to adapt this process to deposit alternating layers of insulating material and high-kinetic-inductance (HKI) material, using a pattern like the one in Fig. 3. This produces a vertical stack of inductors.

Although kinetic inductors do not rely on magnetic fields for energy storage, they are sensitive to strong magnetic fields. If multiple inductors are stacked and carry current in the same direction, the resulting magnetic field may suppress superconductivity and potentially cause the system to crash. To mitigate this effect, adjacent layers may be designed to carry equal currents in opposite directions.

To illustrate, consider a stack of 100 layers based on Fig. 3, separated by insulating layers. Suppose that structures labeled "odd" connect to all odd-numbered layers, and similarly for "even." By connecting the lower odd and even terminals together and using the upper terminals as external connections, adjacent layers carry equal and opposite currents, effectively creating a virtual ground plane between layers.

This configuration yields two inductors, each comprising 50 layers and having 1/50 the inductance of a single-layer structure. Connecting these in series produces a single inductor with 1/25 of the single-layer inductance. Because the inductors are in series, their current is identical in both, with opposing directions in adjacent layers.

The structure effectively consists of 50 conductors in parallel, increasing $I_{max}$ by a factor of 50 while reducing inductance by a factor of 25. Since the maximum stored energy is $E_{max} = ½LI_{max}^2$, the total energy capacity increases by a factor of 100 consistent with scaling expectations.

The structure just described is essentially a stack of folded inductors, where capacitance between the folds would create self-resonance in the device. Electrically, the stack is a parallel connection of the inductors at each layer, so the inductance drops and the capacitance grows with the number of layers—i.e., the number of layers does not change the self-resonance frequency in this idealized model.

The kinetic inductor self-resonance frequency must be considered for the purposes of this document, but if the $f_R << f_{SRF}$, a lumped element model will be sufficient. Per layer, a kinetic inductor wire of length $l_{kiw}$ and width $w$ will have kinetic inductor wire inductance $L_{kiw} = L_□ l_{kiw} / w$ and capacitance $C_{kiw} = (\varepsilon_{diel} / h_{diel}) l_{kiw} \times w$, where $\varepsilon_{diel}$ is the dielectric constant and $h_{diel}$ is the thickness of the dielectric. These yield a lumped-element $f_{SRF} = 1/(2\pi\, l_{kiw}\, \mathrm{sqrt}(L_□ \times \varepsilon_{diel} / h_{diel}))$. Thus, the self-resonant frequency is highest with a short inductor wire and thick insulator.

Furthermore, a simple approximation is that the region area is equal $A_R = l_{kiw} \times w$ and the number of meanders $M = l_{kiw} / \sqrt{A_R}$. Note that HKI processes must use a thin superconducting layer, typically 50 nm or so, but there is no similar constraint on the insulator thickness.

Let us say the initial 3D HKI test chip utilizes an $SiO_2$ insulator thickness $h_{diel}$ = 500 nm $SiO_2$ insulator and wider wires such that the spacing gap is a negligible proportion of the area (so $s$ can be ignored). Based on these assumptions, the SeeQC HKI superconductor layer, and an operating frequency of 1 GHz, the wire width must satisfy $w >> 2.5$ µm. Table 1 contains two examples of about 30 µm, which is within the safe operational region for these boundary conditions.

Such a structure could be fabricated with standard stair-step etching, although a new process that simultaneously contacts all odd or even layers would be advantageous. Details of this process are beyond the scope of this document.

## APPENDIX 3: 4LC SPICE SIMULATION

### A. 4LC Eigenvalues and Eigenvectors

The 4LC circuit (Fig. 5) comprises four identical inductors and four identical capacitances, resulting in eight degrees of freedom. A subsequent appendix provides a rigorous analysis of the circuit, but this appendix shows the results of a SPICE simulation of the simple power supply circuit connected to a reversible logic circuit that converts the four-phase 4LC output into an eight-phase reversible logic circuit. The SPICE script is included in the supplemental material.

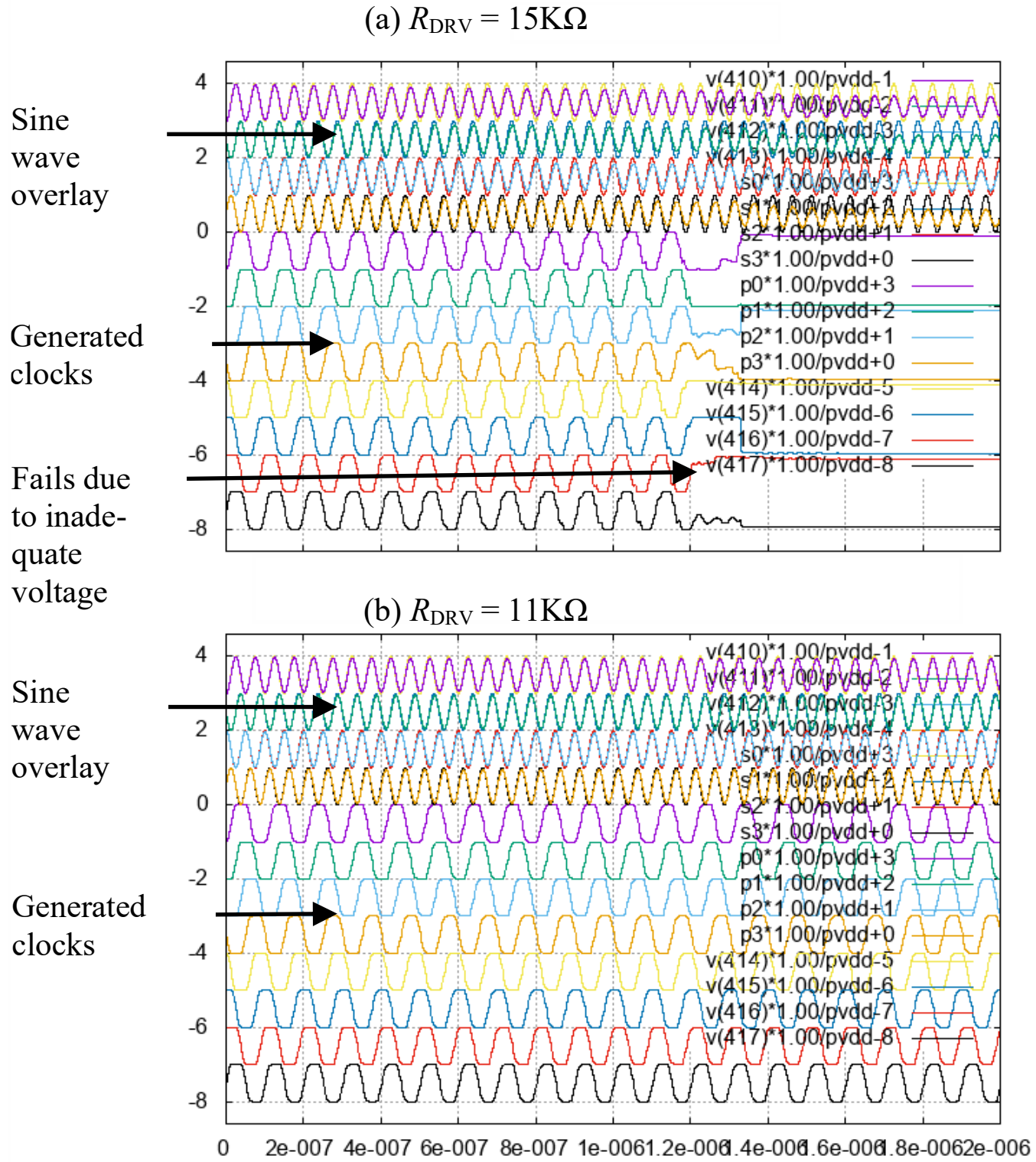

Fig. A1-3. Simulation of 4LC. (a) Works, runs down, and then fails. (b) "Topping off" allows continuous operation.

For clarity, the 4LC circuit does not resonate at the same frequency as a standard LC resonator:

- $f_{LC} = 1 / (2\pi \sqrt{LC})$ (Standard tank)
- $\sqrt{2} \times f_{LC}$ (4LC power-clock mode)
- $2 f_{LC}$ (Parasitic mode)

The $\sqrt{2} \times f_{LC}$ mode produces four sinusoidal waveforms with 90-degree phase separation. Solving the corresponding matrix differential equations yields oscillatory modes in which all four voltages are in quadrature [28]. Small variations in L and C (e.g., a few percent) slightly perturb the phase and frequency but do not lead to cumulative phase drift between clocks.

### B. Demonstration of Sinusoidal Power-Clocks

For each plot in Fig. A3-1, the 4LC circuit was simulated with initial conditions chosen to generate sinusoidal waveforms propagating in the desired direction. The simulations produce four sinusoidal power-clocks that function as the four-phase ramped power-clocks, shown as the upper traces in Fig. A3-1a and Fig. A3-b.

A 2LAL reversible logic circuit generates the phase selectors (Fig. A3-1) labeled as "generated clocks." In Fig. A3-1a, the waveforms are initially well behaved.

## C. Recharging

In the absence of loss, a 4LC circuit would oscillate indefinitely. In practice, losses in transistors and crosstalk cause energy decay.

If component losses are low, the circuit can sustain operation for a finite duration using only the energy provided by the simulation's initial conditions. In the sine wave overlay in Fig. A3-1a, two curves are shown: the 4LC output values $P_n$ and a simulator-generated reference. The power-clock signals decrease over time, eventually reaching a point labeled "fails due to inadequate output voltage," at which point the logic ceases to operate correctly.

Fig. A3-1a and Fig. A3-1b also illustrate a recharging method using resistors $R_{DRV0}$–$R_{DRV3}$. With 15 kΩ resistors (Fig. A3-1a), recharging is insufficient. Reducing the resistance to 11 kΩ in Fig. A3-1b provides adequate energy replenishment, maintaining stable sinusoidal waveforms and consistent multi-phase clock generation.

Alternative implementations may reduce the number of external connections. Consider a single coupling resistor; such a configuration is inherently bidirectional, supporting both leftward- and rightward-propagating waves, which will result in improper operation. This issue can be addressed by using two resistors that preferentially sustain the desired mode while damping undesired modes.

In a further refinement, sinusoidal sources may be replaced with discrete DC voltage levels (e.g., GND, $V_R/2$, and $V_R$). A high-Q resonant circuit requires only intermittent excitation to maintain oscillation. Driving the circuit when the waveform is near these discrete levels can reinforce the desired mode while suppressing others, potentially eliminating the need for continuous analog waveform generation and simplifying implementation.

# Appendix 4: RLC Analysis For 4LC and Reversible Logic

The 4LC circuit and the statistical model of reversible logic circuits both comprise a cycle of four identical subcircuits, which yields a circulant system matrix under an RLC analysis. In this document, an initial "circulant" analysis supports the planning equations in the main text, guides design toward a target clock frequency at the full system power level, and enables co-simulation with the associated logic circuitry. When parameter variance and noise are introduced to individual subcircuits to account for manufacturing variations, transients, and data load, the circulant symmetry breaks. This symmetry breaking splits the degenerate quadrature modes into modes of different frequencies, motivating a new simulation method.

## A. Initial Circulant Analysis

While kinetic inductors were previously presented as ideal, we now introduce a parasitic capacitor ($C_{SRF}$) across their terminals to model device self-resonance.

In this document, kinetic inductor power density serves as a primary metric of scaleup potential. The three-dimensional kinetic-inductor architecture multiplies the energy density by the number of layers; however, parasitic capacitance between adjacent high-kinetic-inductance (HKI) layers, denoted $C_{SRF}$, leads to device self-resonance that limits the maximum clock rate. Increasing the inter-layer dielectric thickness reduces $C_{SRF}$, but the pressure toward higher layer counts and greater vertical separation implies an increase in the total stack height until planarization and fabrication of high-aspect-ratio structures become a scaling limit.

Thus, we define a capacitance variable connecting adjacent clock phases $C_{Adj} = C_{90} + C_{SRF}$.

If we compute the quadrature eigenvalues using the capacitive components $C_{GND}$, $C_{180}$, and $C_{Adj}$, these variables consistently appear in a 1:2:2 ratio. This allows us to simplify the eigenvalue to an expression in $C_{qtr} = C_{GND} + 2C_{180} + 2C_{Adj}$. As mentioned previously, the value of $C_{qtr}$ indirectly defines the value of $\delta_{size}$.

The preceding paragraphs simplify the initial design stages. A designer might be concerned that "crosstalk" capacitance between clock phases would alter the resonant clock rate or energy consumption. Yet, as demonstrated, the designer can reallocate capacitance among the components of $C_{qtr}$ as long as the sum $C_{GND} + 2C_{180} + 2C_{90} + 2C_{SRF}$ remains constant. This simplification enables the planning equations in the main text, but does not apply to the high frequency parasitic mode and, consequently, the impulse response. Furthermore, it does not apply if symmetry is broken through manufacturing variance or data-dependent loading changes. However, these scenarios generally occur later in the design process.

For the integrated inductor architecture presented in this document, all parameters ($C_{GND}$, $C_{180}$, $C_{90}$, and $C_{SRF}$) are approximately proportional to area (e.g., $A_{qtr}$ or the area of a specific region). The parameter $C_{SRF}$ is expected to be relevant only to 3D kinetic inductors and will be proportional to insulator thickness, which is a back-end-of-line parameter. The parameters $C_{GND}$, $C_{180}$, $C_{90}$ are chosen to be dependent on the front-end-of-line semiconductor process and the reversible logic circuit family. Thus, these parameters interact via additive and multiplicative overheads and should be readily manageable.

## B. Reference SPICE Circuit

The circuit described in the main text is defined by the following SPICE netlist:

```
* Inductors (Kinetic)
.param w={2*pi*fR}
L0 P0 P1 {L} IC={-VR/(w*L)} ; iL0 (v0-v1)
L1 P1 P2 {L} IC={-VR/(w*L)} ; iL1 (v1-v2)
L2 P2 P3 {L} IC={VR/(w*L)}  ; iL2 (v2-v3)
L3 P3 P0 {L} IC={VR/(w*L)}  ; iL3 (v3-v0)
* Capacitors (CGND+2*C180 under quadrature assumption)
C0 P0 0 {CGND} IC={0}
C1 P1 0 {CGND} IC={VR}
C2 P2 0 {CGND} IC={0}
C3 P3 0 {CGND} IC={-VR}
C4 P0 P2 {C180} IC={0}
C5 P1 P3 {C180} IC={2*VR}
* Capacitors (Inductor Self-Resonance)
CAdj01 P0 P1 {C90+CSRF}
CAdj12 P1 P2 {C90+CSRF}
CAdj23 P2 P3 {C90+CSRF}
CAdj30 P3 P0 {C90+CSRF}
* Logic Resistance (reversible logic losses)
Rc0 P0 0 {RR}
Rc1 P1 0 {RR}
```

```
Rc2 P2 0 {RR}
Rc3 P3 0 {RR}
* Drive circuitry
Vd0 S0 0 SINE(0 {VR} {fR} 0 0 0)
Vd1 S1 0 SINE(0 {VR} {fR} 0 0 90)
Vd2 S2 0 SINE(0 {VR} {fR} 0 0 180)
Vd3 S3 0 SINE(0 {VR} {fR} 0 0 270)
Rs0 S0 P0 {Rdrv}
Rs1 S1 P1 {Rdrv}
Rs2 S2 P2 {Rdrv}
Rs3 S3 P3 {Rdrv}
```

### C. RLC Analysis Assuming Circulant Symmetry

Let the state vector be defined as $x(t) = [\mathbf{v}(t)^T\ \mathbf{i}_L(t)^T]^T$, where $\mathbf{v}(t) = [v_0\ v_1\ v_2\ v_3]^T$ represents the nodal voltages and $\mathbf{i}_L(t) = [i_{L0}\ i_{L1}\ i_{L2}\ i_{L3}]^T$ denotes the branch currents. The current $i_{Ln}$ is directed from node $P_n$ to $P_{(n+1)\,\mathrm{mod}\,4}$.

Exploiting cyclic symmetry, the system dynamics are decomposed as

$$\mathrm{d}/\mathrm{d}t\, x(t) = Bx(t) + \Gamma \boldsymbol{u}(t),$$

where $B$ is the circulant state matrix and $\Gamma$ is the input distribution matrix that scales the drive voltages.

To construct the state matrix compactly, we define the spatial incidence matrix $M$, which maps the inductive branch currents to the nodes:

$$M = \begin{bmatrix} -1 & 0 & 0 & 1 \\ 1 & -1 & 0 & 0 \\ 0 & 1 & -1 & 0 \\ 0 & 0 & 1 & -1 \end{bmatrix}$$

The inverse capacitance matrix $C^{-1}$ of the symmetric ring is a circulant matrix defined by its first row $[r_0\ r_1\ r_2\ r_1]$, given by the scalar coefficients:

Define $B$ by sub-blocks based on the following scalar coefficients:

$$r_0 = ¼\,[1/C_{\mathrm{GND}} + 2/(C_{\mathrm{GND}}+2C_{\mathrm{Adj}}+2C_{180}) + 1/(C_{\mathrm{GND}}+4C_{\mathrm{Adj}})]$$

$$r_1 = ¼\,[1/C_{\mathrm{GND}} - 1/(C_{\mathrm{GND}}+4\mathrm{C}_{\mathrm{Adj}})]$$

$$r_2 = ¼\,[1/C_{\mathrm{GND}} - 2/(\mathrm{C}_{\mathrm{GND}}+2C_{\mathrm{Adj}}+2C_{180}) + 1/(C_{\mathrm{GND}}+4C_{\mathrm{Adj}})]$$

Let $g = (1/R_{\mathrm{R}} + 1/R_{\mathrm{drv}})$ represent the total conductance to ground, and let $k_l = 1/L$ represent the voltage-to-current scaling. The full state matrix $B$ and input matrix $\Gamma$ are defined via block matrix partitioning:

$$B = \begin{bmatrix} -gC^{-1} & C^{-1}\mathrm{M} \\ -k_l M^T & \mathbf{0}_{4\mathrm{x}4} \end{bmatrix}$$

$$\Gamma = \begin{bmatrix} -1/R_{\mathrm{DRV}}C^{-1} \\ \mathbf{0}_{4\mathrm{x}4} \end{bmatrix}$$

The external drive vector $\boldsymbol{u}(t)$ comprises the four quadrature phases:

$$\mathbf{u}(t) = V_R \begin{bmatrix} \sin(2\pi\, f_R\, t) \\ \sin(2\pi\, f_R\, t + \pi/2) \\ \sin(2\pi\, f_R\, t + \pi) \\ \sin(2\pi\, f_R\, t + 3\pi/2) \end{bmatrix}$$

By defining the exact closed-form expressions for the block system matrix $B$ and input matrix $\mathbf{\Gamma}$, we bypass the need for localized nodal admittance reduction at every simulation step. While computing the exact state at an arbitrary future time analytically would still require the matrix exponential (which relies on eigenvalue decomposition), possessing the explicit block matrices enables highly efficient, rigorous time-step integration using standard ODE solvers without the computational overhead of iterative SPICE-level matrix inversions.

### D. Asymmetry and Data-Induced Capacitance Noise

The simplifications above for nominal operation may be helpful, but we now discuss two forms of off-nominal operation that mandate a more sophisticated mathematical treatment. We will now define the *B'* matrix based on overriding the values of the capacitors specified in the SPICE netlist, specifically the initialization values Cn and CAdjnm are overridden with symbols $C_{\mathrm{n}}$ and $C_{\mathrm{Adjnm}}$.

We define $\mathbf{C}$ as the symmetric 4 × 4 nodal capacitance matrix. To account for individual component variance and device self-resonance across the four phases, the diagonal elements (self-capacitances) are the sum of all capacitors connected to each node:

$$C_{0,0} = C_0 + C_{\mathrm{Adj30}} + C_{\mathrm{Adj01}} + C_4$$

$$C_{1,1} = C_1 + C_{\mathrm{Adj01}} + C_{\mathrm{Adj12}} + C_5$$

$$C_{2,2} = C_2 + C_{\mathrm{Adj12}} + C_{\mathrm{Adj23}} + C_4$$

$$C_{3,3} = C_3 + C_{\mathrm{Adj23}} + C_{\mathrm{Adj30}} + C_5$$

The off-diagonal elements represent the mutual coupling between nodes:

$$C_{0,1} = -C_{\mathrm{Adj01}},\ \ C_{1,2} = -C_{\mathrm{Adj12}}$$

$$C_{2,3} = -C_{\mathrm{Adj23}},\ \ C_{0,3} = -C_{\mathrm{Adj30}}$$

$$C_{0,2} = -C_4,\qquad C_{1,3} = -C_5$$

By symmetry, $C_{\mathrm{i,j}} = C_{\mathrm{j,i}}$

When the individual capacitive components (C0 through C5) vary independently, the system loses its circulant symmetry, which fundamentally alters the spectral properties of the state matrix $B$. In a perfectly symmetric ring, the forward- and reverse-propagating quadrature waves share a degenerate resonant frequency, resulting in identical complex conjugate eigenvalue pairs. Under asymmetric loading, this degeneracy breaks, and the primary quadrature mode splits into two distinct modes with slightly different resonant frequencies and damping rates. In the time domain, this mode splitting manifests as a low-frequency amplitude modulation, or "beating," between the phases. Furthermore, the frequency of the higher-order parasitic mode becomes highly sensitive to mismatches between the cross-phase capacitors C4 and C5. If the $R_{\mathrm{DRV}}$ resistors are small enough, they will correct these distortions. The designer can resolve these issues in simulation using parameter sweeps.

Because exact, closed-form expressions for these split eigenvalues are algebraically intractable under arbitrary component variance, direct time-stepping integration of the explicit *B*' matrix becomes essential. This formulation enables the rigorous evaluation of manufacturing variance and parametric capacitance noise developed in the following section.

Let us further explain the issue by varying only capacitor C5. This capacitor is initialized to $2V_R$ in the netlist and switches polarity twice each clock cycle. To simulate capacitance noise, the simulation must stop every half clock cycle and change C5 to a different capacitance, presumably selected from a distribution around $C_{180}$. The distribution could be based on statistics from measurements of various circuits, or a trace from a logic simulation of the circuit being designed. This change in capacitance must not add or remove energy; therefore the voltage on C5 is adjusted so the post-change energy is equal to the pre-change energy. The simulation then restarts and simulates the next half clock cycle.

Let us define two helper variables:

- $T_{dcy}$ = $1/(R_R\ C_{qtr})$: Time constant for system energy loss.
- $R_{drv} = R_R/10$: Drive stiffness.

Performing this simulation for $C_{GND} = C_{90} = C_{180}$; $C_{SRF} = 0$, $f_R$ = 1 GHz, $V_R$=.11 V, $L$ = 1 nH, $T_{dcy}$ = 20 ns, and an RMS variance of 2 pF around 10.13 pF yields Fig. A4-1.

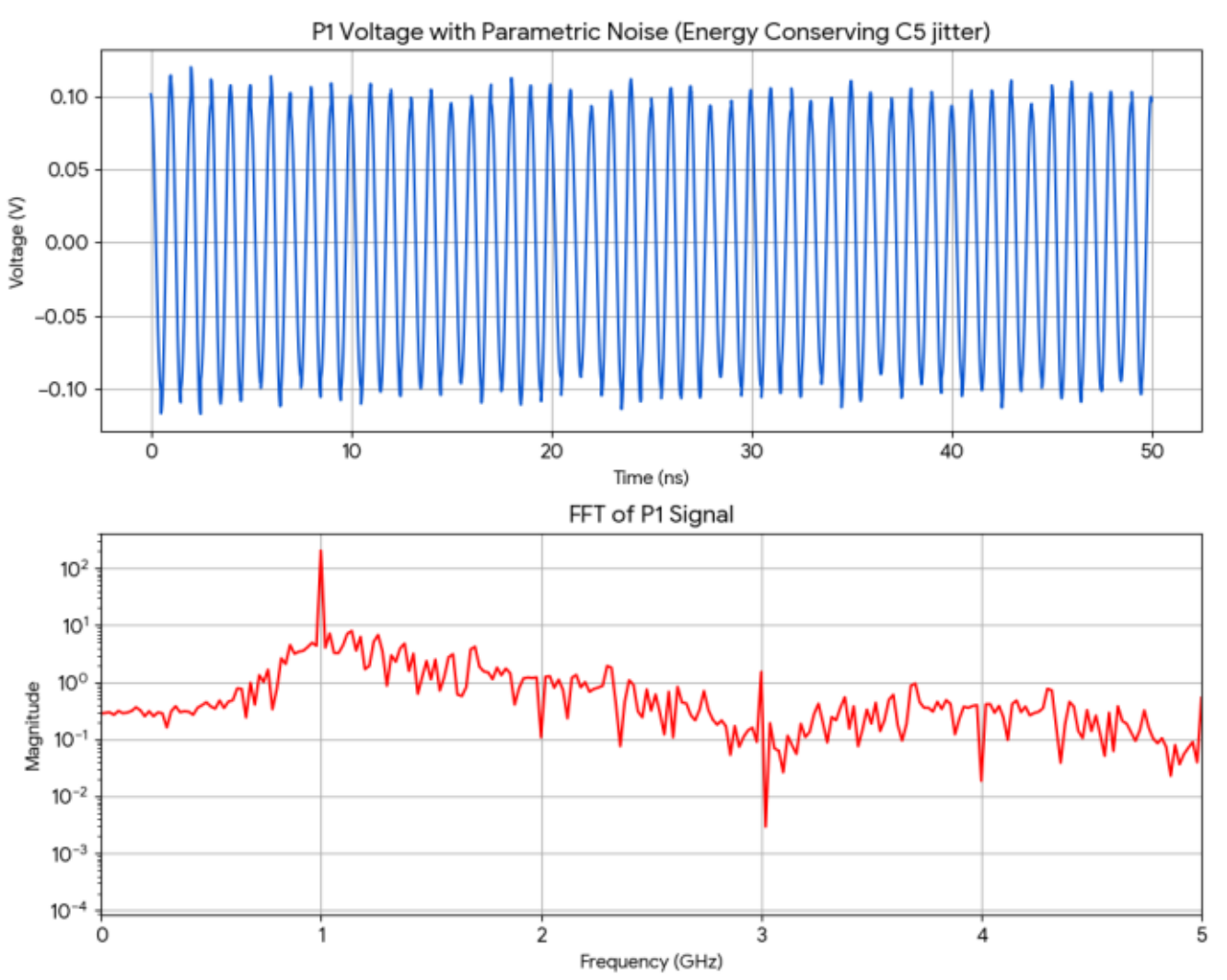

Fig. A4-1: Time domain and spectrum of 4LC system

The charts clearly show the dominance of the 1 GHz clock alongside the variance due to capacitance noise. The capacitance noise breaks the symmetry and excites all the modes.

Extending the noise to all the capacitors defining the reversible logic circuit model (C0-C5), reducing the RMS variance to 0.5 pF and plotting the energy dissipation on each $R_{DRV}$ and $R_R$ yields Fig. A4-2.

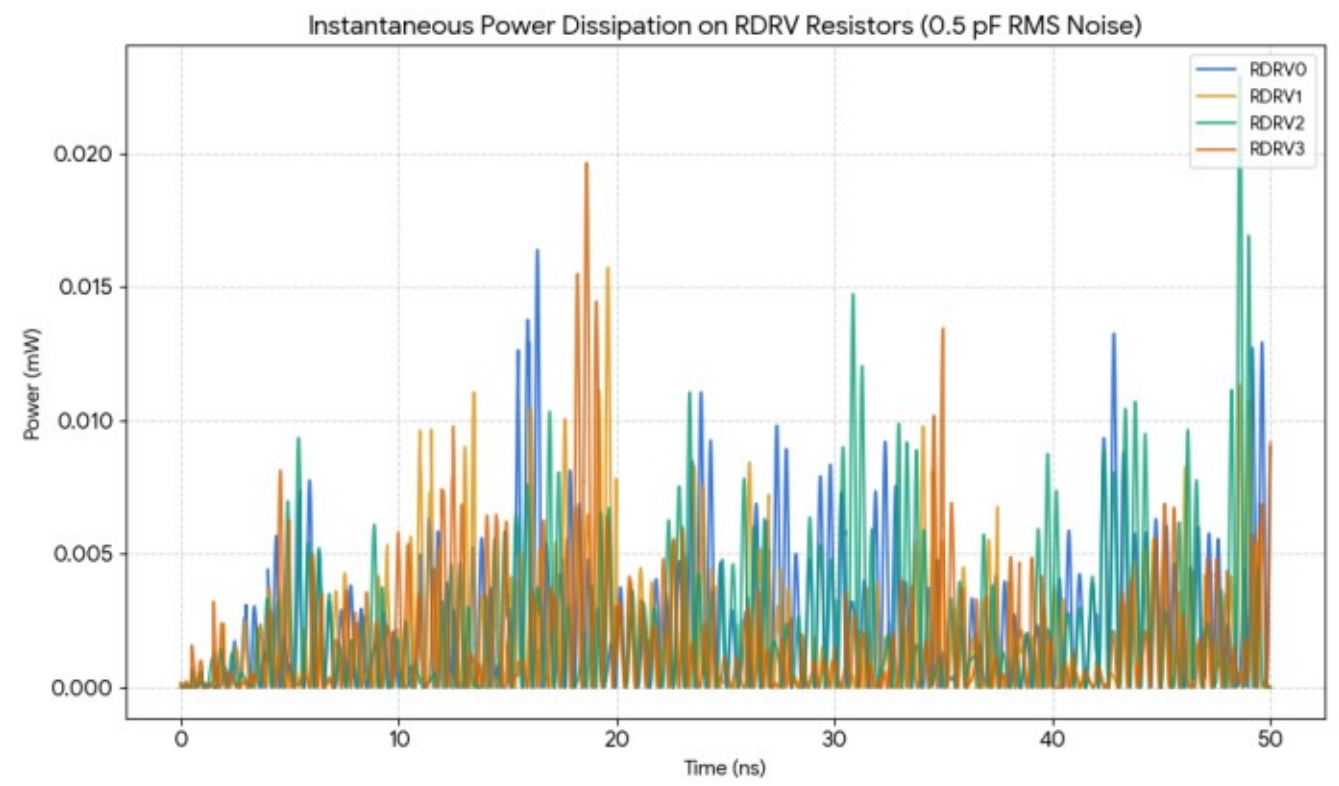

Figure A4-2: Power dissipation of $R_{DRV}$ resistors.

The charts show random bursts, colored either blue-green or orangish. Large capacitance changes cause the larger bursts (such as at 18 ns), which naturally die out over several clock cycles. The blue-green bursts are associated with phases 0 and 2 (the connection points of C4) while the orangish curves are associated with phases 1 and 3 (the connection points of C5).

We now plot the instantaneous and cumulative energy dissipation of $R_{DRV}$ and $R_R$, displaying just one of each to avoid visual clutter, which yields Fig. A4-3.

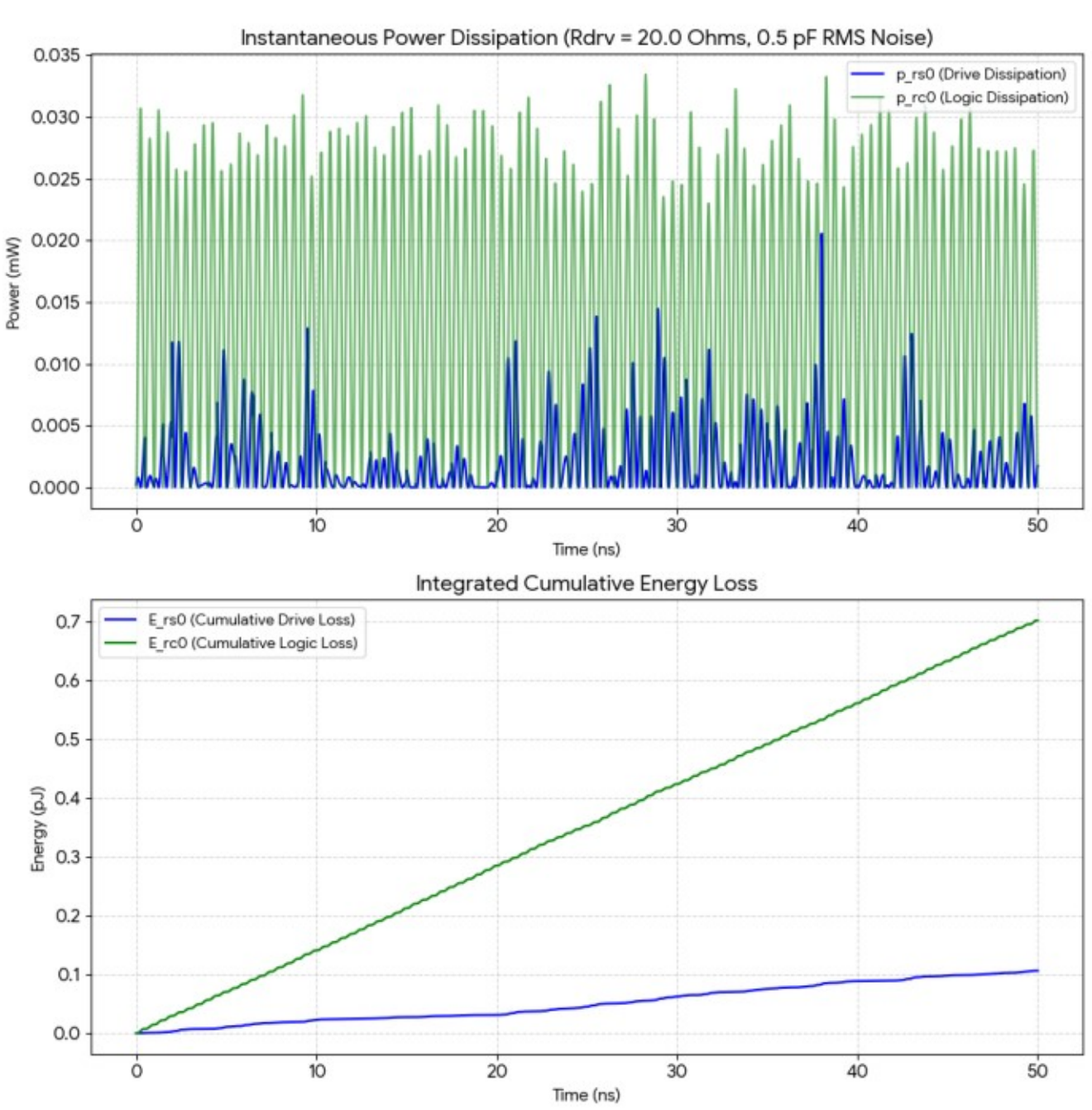

Figure A4-3: Relative power flow of logic vs. $R_{DRV}$.

The green RC0 curve and the nearly-linear envelope in Fig. A4-1 have similar shapes. The blue Rs0 curve is also linear and has about one-seventh the magnitude.

We are now in a position to quantify aspects of the expected energy efficiency improvement derived from energy recycling. In conventional electronics, energy enters a CMOS circuit electrically and leaves as heat. In contrast, energy enters a reversible logic chip electrically and is systematically recycled. Based on this model, the recycling efficiency is defined as the ratio of the power entering $C_C$ divided by the power dissipated in the $R_R$ and $R_{DRV}$ resistors.

We use the following method of computing the energy flowing into the reversible logic chip: The instantaneous power entering/leaving the chip is calculated as the product of the C$N$ voltage and C$N$ current, for all capacitors. Since the chip does not accumulate or lose energy over the long run, the reactive energy delivered to C$N$ during the charging phases is equal to the energy extracted during discharging. Therefore, the total reactive energy entering the capacitor is half the time integral of the absolute value of the instantaneous power flow.

The upper plot shows energy flow into the two types of capacitors in the reversible logic. While the C4 curve indicates a much higher flow than the C0 curve, it is important to note that there are two $C_{180}$ capacitors (C4) but four $C_{90}$ capacitors.

Thus, these figures indicate approximately a 30× improvement in energy efficiency.